\documentclass[lettersize,journal]{IEEEtran}
\usepackage{amsmath,amsfonts}
\usepackage{array}
\usepackage{textcomp}
\usepackage{stfloats}
\usepackage{url}
\usepackage{verbatim}
\usepackage{graphicx}
\usepackage{cite}
\usepackage{todonotes}
\usepackage{xcolor}
\usepackage{subcaption}        
\usepackage[version=4]{mhchem} 
\usepackage[super]{nth}
\usepackage[ruled,vlined,linesnumbered]{algorithm2e} 

\usepackage{soul}

\newcommand\note[1]{\textcolor{brown}{#1}}          
\SetCommentSty{mycommfont}                          

\usepackage{makecell} 
\usepackage{multirow} 

\usepackage{enumitem} 

\begin{document}
\title{Hiding Information for Secure and Covert Data Storage in Commercial ReRAM Chips}


\author{Farah Ferdaus,~\IEEEmembership{Member,~IEEE,}
        B. M. S. Bahar Talukder,~\IEEEmembership{ Member,~IEEE,}
        and Md Tauhidur Rahman,~\IEEEmembership{Senior Member,~IEEE}
\thanks{The authors are with the Department of Electrical and Computer Engineering, Florida International University, Miami, FL 33174 USA (e-mail: fferd006@fiu.edu; bbaha007@fiu.edu; mdtrahma@fiu.edu).}}
\markboth{Journal of \LaTeX\ Class Files,~Vol.~14, No.~8, August~2021}%
{Shell \MakeLowercase{\textit{et al.}}: A Sample Article Using IEEEtran.cls for IEEE Journals}

\IEEEpubid{0000--0000/00\$00.00~\copyright~2021 IEEE}

\maketitle
\begin{abstract}

This article introduces a novel, low-cost technique for hiding data in commercially available resistive-RAM (ReRAM) chips. The data is kept hidden in ReRAM cells by manipulating its analog physical properties through switching (\textit{set/reset}) operations. This hidden data, later, is retrieved by sensing the changes in cells’ physical properties (i.e., \textit{set/reset} time of the memory cells). The proposed system-level hiding technique does not affect the normal memory operations and does not require any hardware modifications. Furthermore, the proposed hiding approach is robust against temperature variations and the aging of the devices through normal read/write operation. The silicon results show that our proposed data hiding technique is acceptably fast with ${\sim}0.4bit/min$ of encoding and ${\sim}15.625bits/s$ of retrieval rates, and the hidden message is unrecoverable without the knowledge of the secret key, which is used to enhance the security of hidden information.

\end{abstract}


\section{Introduction} \label{sec:intro}

\IEEEPARstart{T}{he} most widely used standalone data storage media, Flash, faces enormous performance, scaling, reliability, retention, and cycling challenges in small process nodes. Furthermore, the NAND Flash program and erase operation is slow and can not be performed in small granularity. \textcolor{black}{Therefore, classical security solutions that are elegant and reliable consume a significant amount of energy and can be vulnerable to physical attacks such as radiation and exposure to high temperatures \cite{6691207, 9264740, 7106562}. However, emerging technology-based devices can provide new and robust security primitives and protocols that are potentially stronger than conventional, complementary metal oxide semiconductor (CMOS) device-based security solutions \cite{FRAM_TRNG, MRAM_TRNG}. These emerging devices have inherent process variations and exhibit nonlinear input-output relationships.} To this end, the most promising emerging resistive RAM (ReRAM) technology is gaining attention due to its erase-free simpler architecture, crossbar structure feasibility, fast write performance, lower read latency, more straightforward storage controller design, excellent reliability at high temperature, high endurance, high capacity, high scalability, high performance, ultra-low energy consumption, CMOS compatibility, reduced background memory operations, and reliable storage capability \cite{CrossBar, Fujitsu}. The advantageous features of ReRAM make them the most viable alternative compared to other traditional storage memories \cite{CrossBar, Fujitsu}. The promising aspects of ReRAM make it an ideal candidate to integrate into low-power applications, such as the Internet of Things (IoT), wearable devices (e.g., smartwatches, smart glasses), tablets, smartphones, automobiles, and medical devices (e.g., hearing aids) \cite{CrossBar, Fujitsu}. 

This article demonstrates a technique to embed secret information in a concealed manner in ReRAM cells by leveraging its analog characteristics so that confidential information remains invisible from the normal memory content viewpoint. Our proposed technique can be easily implemented on commercial off-the-shelf (COTS) ReRAM chips, enabling secure and covert data storage. The advantage of the information hiding technique is \textemdash ~ (i) the hidden data can not be retrieved due to watermarking if the storage device is lost or stolen \cite{ReRAM_wMark}, and (ii) the attacker can not read or copy the ciphertext, ensuring an additional protection layer over a typical encryption engine.

Technically, ReRAM is analogous to a two-terminal passive variable resistor where two resistance states, high resistance state ($HRS$) and low resistance state ($LRS$), represent the binary data values. Our technique embeds the hidden information by repeatedly stressing the memory cells by writing ‘1’ and ‘0’ alternatively. Repeated stressing through switching operation (`1’ $\rightarrow$ `0’ or `0’ $\rightarrow$ `1’) gradually decreases the $HRS$ resistance, degrading the memory performance and eventually causing endurance failure \cite{ReRAM_Switching, ReRAM_Mao}. Our experiment indicates that repeatedly stressing the ReRAM cell increases its \textit{write} time (for both logic `0' and `1'). To this extent, we propose a technique of encoding logic `0' and `1' by representing the fresh and stressed memory cells, respectively. Later, we retrieve the encoded sequence by observing the \textit{write} time of corresponding memory cells. Our proposed technique is irreversible as the impact of cell stressing is immutable; hence, the hidden message cannot be tampered with. Additionally, our proposed technique does not require hardware modification and can be directly deployed into available commercial products. Moreover, the embedded message is robust against temperature variation as ReRAM is inherently insensitive to temperature \cite{Bogdan:ReRAM}. Furthermore, our proposed method can be evaluated using standard ReRAM \textit{read/write} operation and only costs ${\sim}3$\% of the total endurance of ReRAM cells. 

\IEEEpubidadjcol
Note that our prior work \cite{ReRAM_wMark} focuses only on watermarking, a traditional application of information hiding, to prevent piracy. Compared to \cite{ReRAM_wMark}, this version presents the information concealing technique, enabling several interesting applications other than watermarking, such as secure and covert data storage, data integrity, and covert communication \cite{6691207}. The key contributions of this article are as follows.

\begin{itemize}[leftmargin=*, topsep=0pt,itemsep=-1ex,partopsep=1ex,parsep=1ex]
    \item {We present a new idea of hiding information in ReRAM by storing logic `0' bit in fresh ReRAM cells and logic `1' in stressed ReRAM cells. We experimentally show that the hidden data can be retrieved by observing ReRAM \textit{write} time.}
    \item {We demonstrate the robustness, performance, retention characteristics, normal memory usage tolerance, and security analysis of our proposed technique in multiple COTS ReRAM chips.}
\end{itemize}

The rest of the paper is organized as follows. Sect. \ref{sec:rel_work} presents the motivation of our work. Sect. \ref{sec:reram} briefly overviews the organization and operating principle of ReRAM chips. 
Sect. \ref{sec:method} presents the proposed information hiding and extracting algorithms. Sect. \ref{sec:result} explains the experimental setup and exhibits the effectiveness and security perspective through obtained results. 
Finally, Sect. \ref{sec:end} concludes the article.

\section{Motivation} \label{sec:rel_work}
This work correlates two research areas \textemdash ~ steganography with hardware security. Due to the rapid evolution in digital multimedia and information technology, hiding information within digital content, e.g., documents, videos, audios, images, text, etc., is gaining significant attention to protect content and intellectual property \cite{Dig_steg, DH_Flash, info_hide, steg, ReRAM_DH}. Prior works on typical digital steganography techniques either use unused meta-data fields or exploit noise in the digital content where the hidden information is usually tied to the digital file itself or the file's content. In \cite{DH_firmware}, firmware defines the hard disk drive's physical locations (sectors), which contains the hidden information, as unusable; hence, the operating system (OS) can not access those sectors, making the recovery process difficult and complicated. {Moreover, the natural aging process introduces significant alterations in the analog domain that change the power-on state of the Static RAM (SRAM) cells, which can also be used for message hiding \cite{DH_SRAM}. Furthermore, exploiting overprovisioning of solid-state drives (SSDs) and Flash Translation Layer (FTL) can be a medium to hide information in the physical layer of an SSD NAND flash memory \cite{deaddrop}. Introducing multiple security levels can avoid sensitive hidden data overwriting while handling data in FTL \cite{DH_SRAM}. However, it is possible to detect, copy, and rewrite the hidden data at a lower security level in the worst case \cite{DEFTL}. In contrast, our proposed scheme depends on intentionally applied stressing to conceal information in inherent analog physical characteristic variations of ReRAM (i.e., \textit{write} time), decoupling the concealed information from the digital memory content and coupling it to physical objects. Retrieving hiding information from physical properties requires detailed and time-consuming measurement and analysis, making detection, copy, and erasure difficult for attackers.   
Therefore, the key benefits of our proposed technique over digital steganography are as follows.

\begin{itemize}
    \item {Our proposed covert technique does not impact normal memory usage or memory content. The hidden information is entirely uncorrelated with the memory content. Therefore, an attacker can not reveal the hidden information by inspecting the memory content or performing memory operations. Our experimental results reveal that the hidden information remains intact even after thousands of normal memory operations.}
    \item {Unlike digital steganography, hidden information can not be copied or stored for future analysis as our technique manipulates analog physical characteristics. For example, suppose an attacker gains access to the ReRAM without knowledge of the location of the hidden bit. In that case, it will not be possible to reveal confidential information by only measuring the \textit{set/reset} time of the memory cells. Performing a brute force attack is not feasible as it requires a vast amount of time.}
\end{itemize}

Besides, other steganographic methods modify the physical layer protocol to hide data in the noise of optical and wireless transmission instead of encoding in digital objects \cite{steg_optical, steg_ofdm, steg_wireless}. These techniques require either special tools or modifications to existing protocols. In contrast, our proposed technique is analogous to steganography, where a physical object or digital information hides a confidential message \cite{Phy_Steg, DH_Flash, ReRAM_DH}. The key benefit of our proposed cost-effective ReRAM-based technique is \textemdash ~ the embedding/retrieval of the concealed information can be performed without any circuit/hardware modification or subject-matter experts \cite{Watermarking}. Furthermore, our easily applicable scheme can be implemented using a software program in a low-level memory interface and works with standard digital interfaces with COTS memory chips.


\section{ReRAM Preliminaries} \label{sec:reram}

Resistive switching phenomena in a dielectric material is the core mechanism of ReRAM to store logic states \cite{ReRAM_hist, ReRAM_Mao}. The capacitor-like ReRAM cell structure consists of two electrodes ($Electrode_{Top}$ and $Electrode_{Bottom}$) separated by a metal oxide resistive switch material (Fig.~\ref{fig:ReRAM}). Studies show that various metal oxide materials can be used to build the resistive switch layer, such as $\ce{Al2O3}$, $\ce{NiO}$, $\ce{SiO2}$, $\ce{Ta2O5}$, $\ce{ZrO2}$, $\ce{TiO2}$, $\ce{HfO2}$, and $\ce{Nb2O5}$ \cite{ReRAM_hist, ReRAM_Mao}. However, different materials result in different device characteristics such as endurance, retention, and scalability \cite{ReRAM_hist, ReRAM_Mao}. 
Whenever a voltage is applied to the $Electrode_{Top}$, the metal oxide breakdown process is initiated, producing oxygen vacancies in the oxide layer. Consequently, these oxygen vacancies form a conductive filament between two electrodes and produce the low resistance state ($LRS$ or logic `0' state). A voltage with opposite polarity is applied across the metal oxide to eliminate the conductive filament, representing the high resistance state ($HRS$ or logic `1' state) of the ReRAM cell. The ratio of $HRS$'s resistance to $LRS$'s must be large enough to ensure robust \textit{read/write} operation \cite{ReRAM_Mao}. The switching operations from $HRS$ ($LRS$) to $LRS$ ($HRS$) is known as \textit{set} (\textit{reset}) operation, and the time required for switching is known as the \textit{set} (\textit{reset}) time. 
In summary, the ReRAM \textit{read/write} operation is performed as follows-
\begin{itemize} 
    \item {The \textit{write} operation ensures appropriate voltage magnitude and polarity across the ReRAM cell; as a result, the ReRAM cell obtains the appropriate resistance state ($LRS$ for logic `0' and $HRS$ for logic `1').} 
    \item {During the \textit{read} operation, a small voltage is applied across the ReRAM bit cell, and the measured resistance (by sensing current) determines the stored logic state.}
\end{itemize}

\begin{figure}[ht!]
    \centering
    \captionsetup{justification=centering, margin= 0cm}
    \includegraphics[trim=0cm 13.1cm 16.8cm 2.8cm, clip, width = 0.32\textwidth]{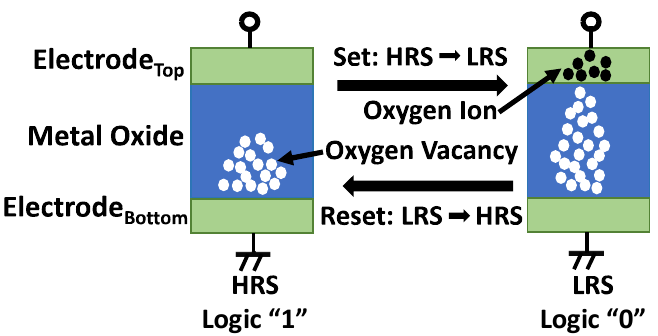}
    \caption{ReRAM cell structure with two logic states \cite{ReRAM_Mao}.}
    \label{fig:ReRAM}
\end{figure}

Each switching operation (i.e., changing state from $LRS \rightarrow HRS$ or $HRS \rightarrow LRS$) on ReRAM gradually decreases the resistance of $HRS$, wearing-out the device \cite{ReRAM_Switching}. Hence, fresh memory cells possess distinctly different analog properties from stressed cells (i.e., cells that undergo repeated switching operations). For example, the reduction of resistance of $HRS$ due to the wear-out process degrades the resistance ratio ($\frac{R_{HRS}}{R_{LRS}}$) \cite{ReRAM_Switching, ReRAM_Mao}. To maintain the desired resistance ratio, 
\textit{set} and \textit{reset} times must be increased for stressed memory cells\footnote{The ReRAM internal control circuit maintains appropriate \textit{set/reset} time by initiating the write-verify-write operation sequence \cite{ReRAM_wMark}.}. This work uses this property to distinguish between the fresh and stressed ReRAM cells. 

\section{Proposed Hiding Technique} \label{sec:method}

This section describes the concealing and retrieval techniques of our information-hiding scheme, along with the cell characterization performed to understand the analog physical characteristics of ReRAM cells.

\subsection{Cell Characterization} \label{subsec:Char}

Repeated switching operations (alternatively writing 0's and 1's) change the physical properties of ReRAM; therefore, the \textit{set/reset} timing of stressed cells deviates from the fresh cells. The degree of deviation depends on the number of switching operations performed on stressed cells. Our proposed technique imprints logic `1’ with stressed cells and `0’ with fresh cells.  Later, we retrieve the data by separating the fresh and stressed cells based on their switching time. However, ReRAM stressing reduces cell endurance. Therefore, we want to keep the stress level as little as possible and simultaneously ensure that fresh and stressed cells are reliably separable with \textit{set/reset} time.

To this extent, we examine the ReRAM cell characteristics, and the impact of switching operations on \textit{set/reset} timing. This allows us to determine the minimum number of switching operations required to reliably separate the stressed cell from the fresh cell. It also builds a relationship between ReRAM switching time and corresponding stressing level. To this end, we write all `1’ data patterns to selected memory addresses. Then, all `0' and all `1' data patterns are written alternatively to those addresses. The switching times are captured and stored as \textit{set/reset} times accordingly. We repeat the switching operation until the target memory cells are fully worn out (i.e., no longer able to store data reliably). We observe that both the \textit{set} and \textit{reset} times increase due to the repeated switching operation, and after a certain number of switching operations, the stressed cells completely become separable from fresh cells.

Note that, according to our observation, the relation between switching characteristics (i.e., \textit{set/reset} time vs. stress count\footnote{One `stress' means a pair of \textit{set-reset} operations.}) is almost uniform for all memory chips sharing the same part-number. Therefore, it should be sufficient to sample a small set of memory chips from each part-number and perform cell characterization over those chips.

\subsection{Information hiding Technique} \label{subsec:encode}

Our proposed technique conceals information in the \textit{write} (\textit{set/reset}) time of ReRAM bit cells. The \textit{set} and \textit{reset} time increases monotonically with stress levels, making it possible to retain a hidden message and retrieve it through a proper threshold value that separates the stressed and fresh memory cells \cite{ReRAM_wMark}. The authorized entity performs the proposed information concealing technique in the memory to encode secret information in ReRAM. In the proposed technique, we choose a set of addresses for the secret message; the number of addresses depends on the length of the secret message. Initially, all memory cells possess perfect or near-perfect analog properties since they are fresh. To encode a secret message, (i) initially, logic `1' is written to those chosen addresses (line \ref{alg2:init0} through line \ref{alg2:init1} of Algorithm \ref{alg:dataHide}), and (ii) repeated switching (\textit{set} and \textit{reset}) operations are performed (line \ref{alg2:dHide0} through line \ref{alg2:dHide1} of Algorithm \ref{alg:dataHide}) to only those ReRAM addresses, which are supposed to hold the logic `1' of the secret message. The switching operations are repeated until sufficient differences are developed in the \textit{set/reset} time between fresh and stressed memory cells. Each switching operation gradually degrades the resistance of $HRS$, which is permanent and, thus, cannot be reversed. However, the number of repeated switching cycles, $\mathcal{N}$, used to encode the secret message is determined empirically through cell characterization for given memory chips to ensure proper separation without causing excessive stress \cite{ReRAM_wMark}. From an encoding perspective, minimizing $\mathcal{N}$ is desirable because the encoding time of the secret message is directly proportional to the number of switching cycles. However, higher $\mathcal{N}$ enhances the accuracy by distinguishing fresh and stressed memory cells more perfectly even after thousands of memory operations. The silicon results show that a few thousand rewrite operations are sufficient to hide information securely.

\begin{algorithm}[ht!]
\SetAlgoLined
    \KwData{
        $\mathcal{N}$: \note{Number of stress count (i.e. \textit{set-reset} pairs)} \break 
        $\mathcal{A_M}$: \note{Set of memory addresses containing secret message.} \break
        $w_L$: \note{Word Length} \break
        $\mathcal{S}_{msg}$: \note{Secret message} \break
        $\mathcal{D}$: \note{($1 \times w_L$) matrix containing data intended to write each memory cells} \break
        $t$: \note{Timer}} \break
            
    \KwResult{
        $\mathcal{S_T}$: \note{\textit{Set} time of each memory address belongs to $\mathcal{A_M}$} \break
        $\mathcal{R_T}$: \note{\textit{Reset} time of each memory address belongs to $\mathcal{A_M}$}}
    
    \BlankLine    
    \tcp{Initialization} 
    $\mathcal{S_T} = \{ \};\ \mathcal{R_T} = \{ \};\ \mathcal{D} = Ones(1 \times w_L)$\;
    \ForEach{a $\in \mathcal{A_M}$}{ \label{alg2:init0}
        $write(a, \mathcal{D})$\;
    } \label{alg2:init1}

    \BlankLine
    \tcp{Encoding secret message}
    \For{$i = 0$ to $\mathcal{N}$}{ \label{alg2:dHide0}
        \ForEach{a $\in \mathcal{A_M}$}{ 
            \If{$\mathcal{S}_{msg}$[Bit]==1}{
                $\mathcal{D} = Zeros(1 \times wL)$\;
                $write(a, \mathcal{D})$\;
                $\mathcal{D} = Ones(1 \times wL)$\;
                $write(a, \mathcal{D})$\;
            } 
        }
    } \label{alg2:dHide1}

    \BlankLine
    \tcp{Decoding secret message}
    \ForEach{a $\in \mathcal{A_M}$}{ \label{alg2:read0}
        $\mathcal{D} = Zeros(1 \times w_L)$\;
        $tic = t$\;
        $write(a, \mathcal{D})$; \tcp{\textit{Set} operation} \break
        $toc = t - tic$; \tcp{Accumulating \textit{Set} time} \break
        $\mathcal{S_T} = \mathcal{S_T} \cup \{toc\}$\; 
        \BlankLine
        $\mathcal{D} = Ones(1 \times w_L)$\;
        $tic = t$\;
        $write(a, \mathcal{D})$; \tcp{\textit{Reset} operation} \break
        $toc = t - tic$; \tcp{Accumulating \textit{Reset} time} \break 
        $\mathcal{R_T} = \mathcal{R_T} \cup \{toc\}$\;
    }\label{alg2:read1}
 \caption{Pseudo-code for encoding and decoding secret message.}
 \label{alg:dataHide}
\end{algorithm}

\subsection{Hidden Information Retrieval Technique} \label{subsec:retrive}

In order to retrieve concealed information, the physical properties of memory cells are extracted (in our case, \textit{set/reset} times) to distinguish between fresh and stressed memory cells. Line \ref{alg2:read0} to \ref{alg2:read1} of Algorithm \ref{alg:dataHide} outlines the required steps for extracting the \textit{set} and \textit{reset} times from the addresses containing concealed information. \textcolor{black}{In line 19, we accumulate all the \textit{set/reset} times in a single `bag' (or set) from the addresses containing hidden information. Therefore, if the imprinted value is 32-bit long, we will have 32 individual \textit{set/reset} times in the bag.} We observe that both \textit{set} and \textit{reset} time change with stress counts, and both can be used to hide the secret message. In practice, we observe an apparent gap between the average write times of the fresh and stressed memory cell clusters with sufficient switching operations. As a result, it is straightforward to define the threshold value of \textit{set/reset} time after encoding the secret message, which can be used to differentiate between fresh and stressed memory cell clusters. \textcolor{black}{Based on the threshold value of \textit{set/reset} time, we identify the logic level (logic `0' or `1') of the memory cells that are bagged previously in line 19.} It is worth mentioning that \textit{set/reset} characteristics of ReRAM cells appear to be uniform across all ReRAM chips that we have tested.

\subsection{Encoding/Decoding Secret Information} \label{subsec:AddtionalSecurity}
The flowchart in Fig. \ref{fig:dataHide_step} shows the steps of the information hiding process, i.e., chronologically encoding and decoding information in ReRAM. The information-hiding operation steps are pretty straightforward. We hide the secret message in analog physical characteristics of ReRAM by repeatedly stressing the memory cells. To this end, first, we characterize a few memory cells to understand the analog physical characteristics of ReRAM cells at different stressing levels up to the maximum endurance \cite{ReRAM_wMark}. Later, the authorized personnel performs the retrieval operation to extract the hidden message bits from the analog physical characteristics through standard digital interfaces when required. The initial address of the stored information, replica size, and the shift sequence of each hidden bit are used as a hiding key during both hiding and recovery operations. Adding the error-correcting code (ECC)\footnote{Detail ECC algorithm is out of this paper’s scope.} to the secret message can improve the information recovery process by correcting obtained bit errors.

\begin{figure}[ht!]
    \centering
    \captionsetup{justification=centering, margin= 0cm}
    \includegraphics[trim=0cm 11.6cm 22.5cm 0cm, clip, width = 0.3\textwidth]{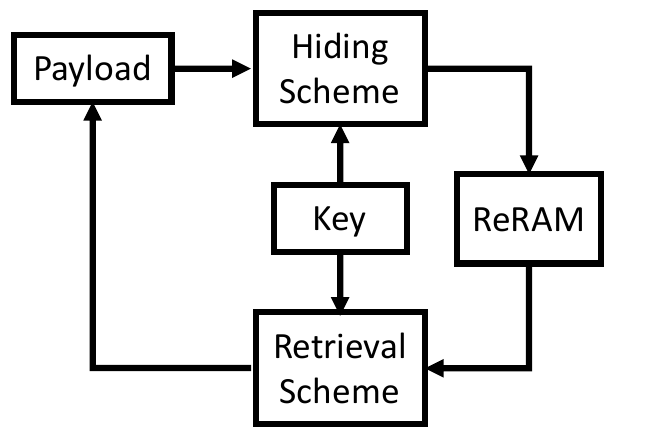}
    \caption{Operation steps used to hide information.}
    \label{fig:dataHide_step}
\end{figure}

We assume that an attacker gets temporary access to the memory that contains hidden information so that they can check, inspect, and manipulate the hidden information through normal memory operations using a standard digital interface. We also assume that the attacker is aware of the information-hiding technique so that they can also examine the analog physical characteristics through the standard digital interface. Therefore, our target is to develop the hiding scheme so that it takes a sufficiently long time and effort to detect the existence and retrieve or remove the hidden information.

\section{Experimental Results} \label{sec:result}

In this section, we evaluate our proposed technique through experiments on COTS ReRAM chips. In addition to validating the correct operation of the information hiding and retrieval scheme, we also analyze the robustness, performance, retention characteristics, recovery without the hiding key, and normal memory usage tolerance.

\subsection{Evaluation Setup}

The analysis is performed over five \textit{MB85AS8MT} (40nm technology node) 8-bit serial peripheral interfaced (SPI) $8Mb$ memory chips manufactured by Fujitsu Semiconductor Limited. We have used our own custom-designed memory controller implemented on \textit{Teensy 4.1} microcontroller development board to issue commands and receive data from the memory chip. These multiple-chip samples are used to verify the proposed technique and determine its feasibility. The \textit{MB85AS8MT} ReRAM chips are byte-addressable. Therefore, a single byte is the smallest unit for which we can measure \textit{set/reset} time. As a result, we need at least a one-byte storage area in the ReRAM to encode a single bit of data. However, the measured \textit{set/reset} time might vary due to the external and internal noise. Therefore, for simplicity, we encode single-bit data into 256 consecutive addresses of the ReRAM to suppress the impact of noise (i.e., single-bit is replicated over 256 addresses). Sect. \ref{subsec:grpSz} discusses the impact of using different replica sizes. Moreover, it is possible to encode data into non-consecutive addresses to make the detection scheme more complex (discussed in Sect. \ref{subsubsec:inac_hideKey}). Therefore, to increase the complexity of the retrieval process, instead of encoding single-bit data into 256 consecutive addresses, we can use any stream cipher (\cite{cusick2004stream}) to select the memory addresses for each message bit. The memory addresses and replica selections are based on the ``hiding key" (Fig. \ref{fig:dataHide_step}) in a way that cannot be predicted without the key.

We have measured the \textit{set/reset} time for each address and computed the average for the evaluation. From now on to the rest of the paper, we denote the average \textit{set/reset} time over 256 addresses as $t_{Set,256}$, and $t_{Reset,256}$, respectively. Note that the \textit{write buffer} size of our tested ReRAMs is also 256, which enables us to stress 256 addresses with a single \textit{write} command, reducing overall stressing time. Although the figures presented in this section are based on a single ReRAM chip (randomly chosen from five test chips), the observation is valid for all test chips. Additionally, Tab. \ref{Tab:summ} summarizes the post-hiding stress tolerance results from all five test chips.

The following steps are performed to verify the feasibility of the proposed information-hiding technique. We have embedded an arbitrarily chosen 32-bit random data (0xECE3038B\footnote{Also verified for other random data.}) into $(256 \times 32) = 8192$ memory addresses, varying the number of switching cycles, $\mathcal{N}$, up to $45K$ times to demonstrate the information hiding (discussed in Sect. \ref{subsec:encode}) and retrieval (discussed in Sect. \ref{subsec:retrive}) technique experimentally. To study whether the proposed technique can reliably hide and recover bits in the \textit{write} (\textit{set/reset}) time characteristics, we use the separation between logic `0' and `1' as the evaluation metric. 


\subsection{Cell Characterization} \label{subsec:cell_char}

The switching characteristics (\textit{set/reset} time vs. the stress counts) of the ReRAM chips at $25^{\circ}C$ are shown in Fig. \ref{fig:char}. These figures represent the maximum, minimum, and average of $t_{Set,256}$ (Fig. \ref{fig:SetT_char}) and $t_{Reset,256}$ (Fig. \ref{fig:ResetT_char}) as a function of different stress levels (up to maximum possible rewrite operations\footnote{Maximum rated endurance for \textit{MB85AS8MT} ReRAM chip is $1M$ rewrite cycles (i.e., $500K$ \textit{set-reset} pairs). However, we observe that most memory cells can endure more rewrite operations than the rated endurance. In our experiment, we stress memory cells with up to $1M$ \textit{set-reset} pairs.}) over the 2K random address space. Fig. \ref{fig:char} demonstrates that both the $t_{Set,256}$ and $t_{Reset,256}$ increase monotonically with stress levels, making it possible to distinguish between stressed and fresh memory cells. For example, the right-side zoomed plot of Fig. \ref{fig:SetT_char}, and Fig. \ref{fig:ResetT_char} represents \textit{set/reset} time up to $50K$ stress count, which demonstrates that the minimum value of $t_{Set,256}$ and $t_{Reset,256}$ at stressed count ${\sim}12K$ is larger than the maximum value of $t_{Set,256}$ and $t_{Reset,256}$ at fresh condition. Therefore, a proper threshold value of $t_{Set,256}$ or $t_{Reset,256}$ can reliably identify fresh and stressed cells with ${\sim}12K$ \textit{set/reset} operations. Although Fig. \ref{fig:char} is constructed with $2K$ memory addresses, a similar characteristic is valid for the whole address space.

\begin{figure}[ht!]
    \centering
    \captionsetup{justification=centering, margin= 0.5cm}
    \begin{subfigure}[t]{0.49\textwidth}
        \centering
        \includegraphics[trim=0.2cm 9cm 4.5cm 0.1cm, clip, width=0.9\textwidth]{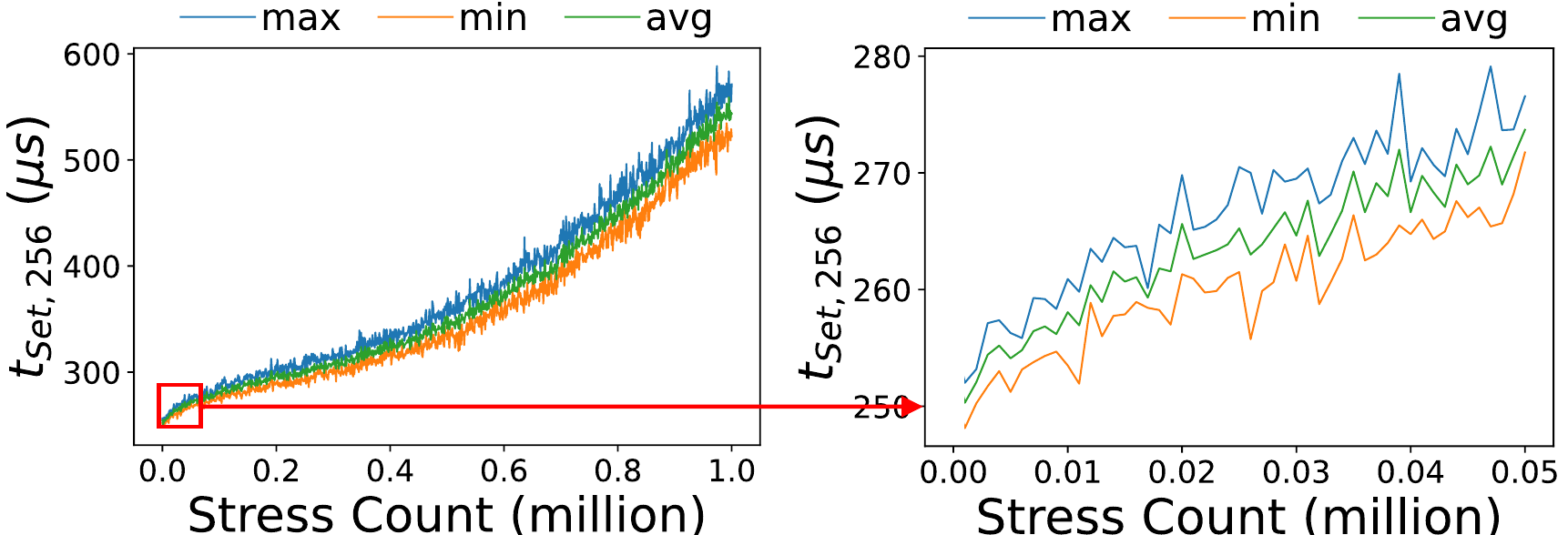}
        \caption{}
        \label{fig:SetT_char}
    \end{subfigure}%
    \vspace{\medskipamount}
    \begin{subfigure}[t]{0.485\textwidth}
        \centering
        \includegraphics[trim=0.2cm 8.9cm 4.4cm 0cm, clip, width = 0.9\textwidth]{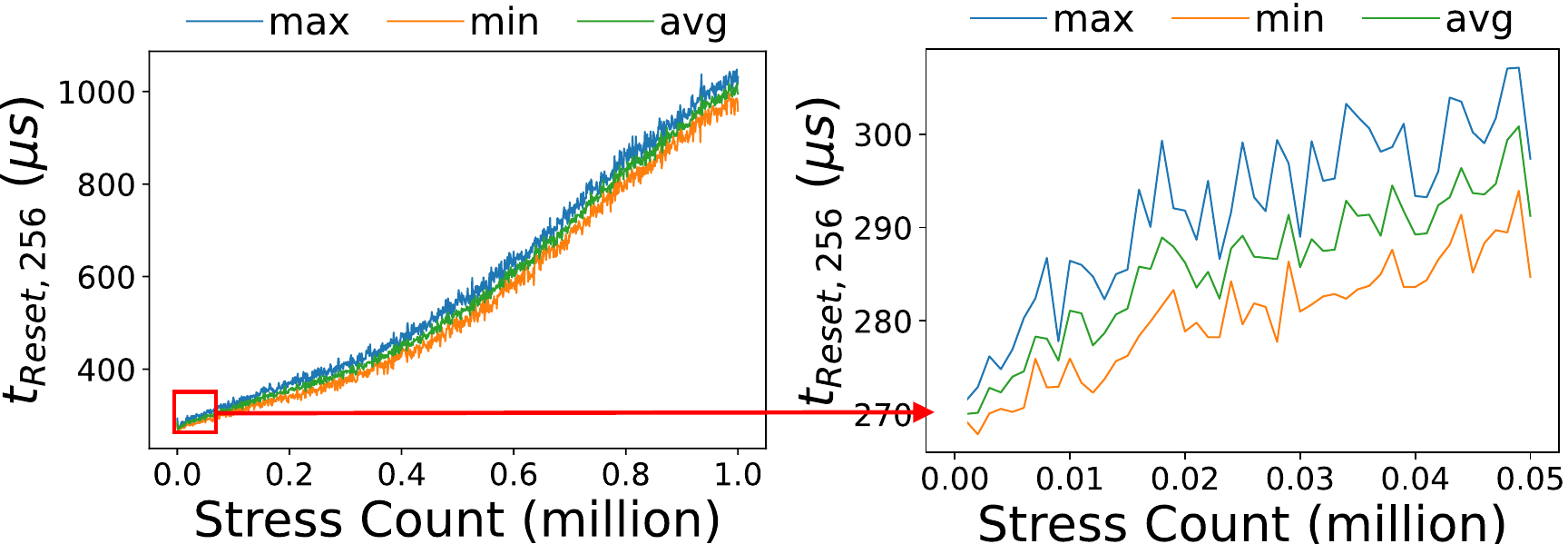}
        \caption{}
        \label{fig:ResetT_char}
    \end{subfigure}
    \caption{ReRAM cell characterization under stress-\\ (a) $t_{Set,256}$ and (b) $t_{Reset,256}$.}
    \label{fig:char}
\end{figure}
%

\subsection{Appropriate Replica Size Selection} \label{subsec:grpSz}

Fig. \ref{fig:summary_grSz} illustrates the impact of replica size (i.e., consecutive addresses used to encode single-bit data). This figure represents the distribution of $d(b_0,b_1)$ at different replica sizes, where $d(b_0,b_1)$ represents the distance between logic `0' bits ($b_0$) and logic `1' bits ($b_1$). Each dot in Fig. \ref{fig:summary_grSz} represents $d(b_0^i,b_1^j)$ for each possible $(b_0^i,b_1^j)$. The figure demonstrates that the separation between logic `0' bits and logic `1' bits improves with respect to replica size. \textcolor{black}{While using \textit{set} time, the bit `0' and bit `1' are well separable  (i.e., $d(b_0,b_1) > 0$, for all (i, j)) with the smallest possible replica size of 32 (Fig. \ref{fig:summary_grSz_Set}). On the other hand, \textit{reset} time only distinguishes the bit `0' and `1' with the minimum replica size of 224 (Fig. \ref{fig:summary_grSz_Reset}). Therefore, in our proposed data hiding technique, we can use any value above 32 while using \textit{set} time and any value above 224 while using \textit{reset} time.}
A smaller replica size will increase the performance by reducing the imprinting and retrieval time. In our implementation, we have chosen the replica size of 256 for the following two reasons-
\begin{itemize}[leftmargin=*, topsep=0pt,itemsep=-1ex,partopsep=1ex,parsep=1ex]
    \item \textcolor{black}{For both $t_{Set}$ and $t_{Reset}$, $d(b_0,b_1) > 0$ with the replica size of 256.}
    \item \textcolor{black}{All of our test ReRAM chips have an internal data buffer that can hold data for up to 256 addresses. This buffer is an unmodifiable hardware component (a set of 8-bit registers). These ReRAMs also come with dedicated instructions, enabling us to send data for 256 addresses from the memory controller to the memory chip simultaneously and write them in parallel. Therefore, choosing 256 replica sizes simplified the experimental setup and reduced the implementation complexity. }
\end{itemize}

 

\begin{figure}[ht!]
    \centering
    \captionsetup{justification=centering, margin= 0.5cm}
    \begin{subfigure}[t]{0.227\textwidth}
        \centering
        \includegraphics[trim=0.1cm 0.1cm 0.1cm 0.1cm, clip, width = 0.9\textwidth]{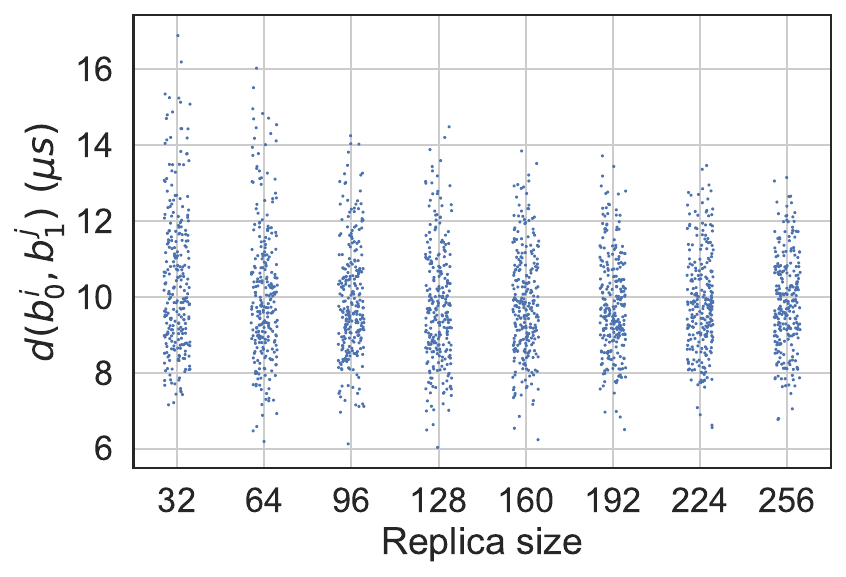}
        \caption{}
        \label{fig:summary_grSz_Set}
    \end{subfigure}
    \begin{subfigure}[t]{0.235\textwidth}
        \centering
        \includegraphics[trim=0.1cm 0.1cm 0.1cm 0.1cm, clip, width = 0.9\textwidth]{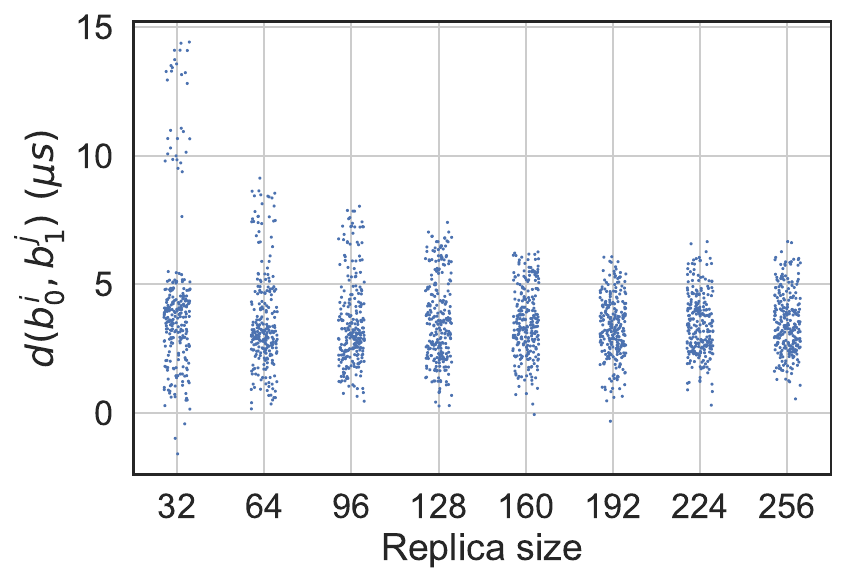}
        \caption{}
        \label{fig:summary_grSz_Reset}
    \end{subfigure}
    \caption{Influence of replica size on the hidden information, using- (a) $t_{Set,[32,256]}$, and (b) $t_{Reset,[32,256]}$.}
    \label{fig:summary_grSz}
\end{figure}

\subsection{Retention Characteristics}

The retention characteristics of the proposed hiding technique are also studied and shown in Fig. \ref{fig:retention_Set} and Fig. \ref{fig:retention_Reset}. Note that since each retrieval performs one \textit{set-reset} operation, these retention characteristics include impacts from additional \textit{set-reset} operations in addition to the time between information hiding and retrieval. The red and blue dots in Fig. \ref{fig:retention_Set} and Fig. \ref{fig:retention_Reset} represent the imprinted logic 1's and 0's, respectively. Fig. \ref{fig:retention_init_Set} and Fig. \ref{fig:retention_init_Reset} show that logic `1' and logic `0' are well separated at stress level $15K$, which is used to hide information considering both $t_{Set,256}$ and $t_{Reset,256}$, respectively. After over two months of retention, the logic `1' and logic `0' remain separated for both $t_{Set,256}$ (Fig. \ref{fig:retention_2m_Set}) and $t_{Reset,256}$ (Fig. \ref{fig:retention_2m_Reset}). The results verify that the retention time has little or no impact on bit separation. 

\begin{figure}[ht!]
    \centering
    \captionsetup{justification=centering, margin= 0.5cm}
    \begin{subfigure}[t]{0.227\textwidth}
        \centering
        \includegraphics[trim=0.2cm 0.3cm 0.2cm 0.5cm, clip, width = 0.9\textwidth]{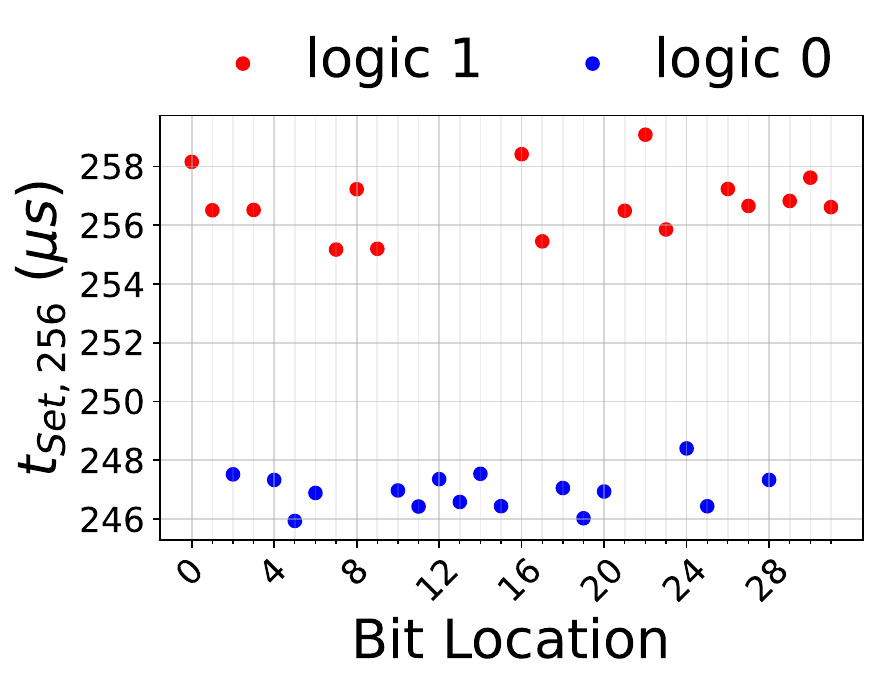}
        \caption{Zero retention.}
        \label{fig:retention_init_Set}
    \end{subfigure}
    \begin{subfigure}[t]{0.235\textwidth}
        \centering
        \includegraphics[trim=0.2cm 0.3cm 0.2cm 0.5cm, clip, width = 0.9\textwidth]{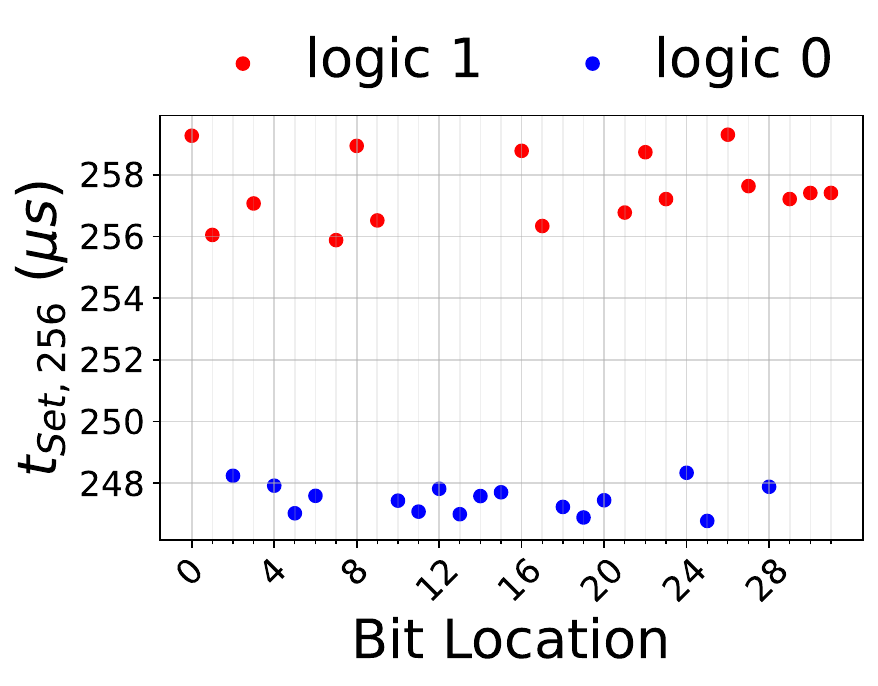}
        \caption{Over two months retention.}
        \label{fig:retention_2m_Set}
    \end{subfigure}
    \caption{Retention characteristics of the hidden information using $t_{Set,256}$.}
    \label{fig:retention_Set}
\end{figure}

\begin{figure}[ht!]
    \centering
    \captionsetup{justification=centering, margin= 0.5cm}
    \begin{subfigure}[t]{0.227\textwidth}
        \centering
        \includegraphics[trim=0.2cm 0.3cm 0.2cm 0.5cm, clip, width = 0.9\textwidth]{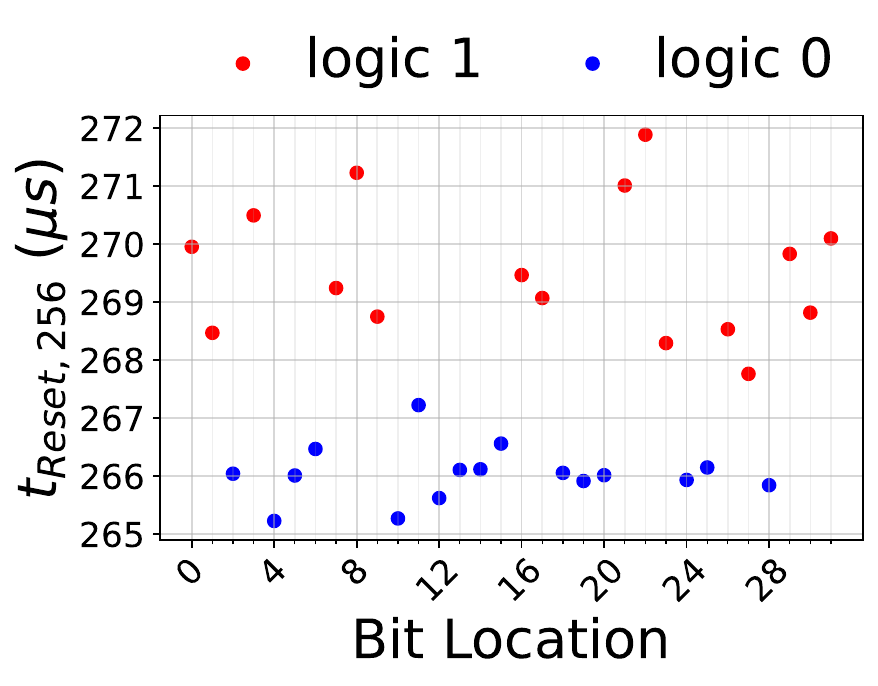}
        \caption{Zero retention.}
        \label{fig:retention_init_Reset}
    \end{subfigure}
    \begin{subfigure}[t]{0.235\textwidth}
        \centering
        \includegraphics[trim=0.2cm 0.3cm 0.2cm 0.5cm, clip, width = 0.9\textwidth]{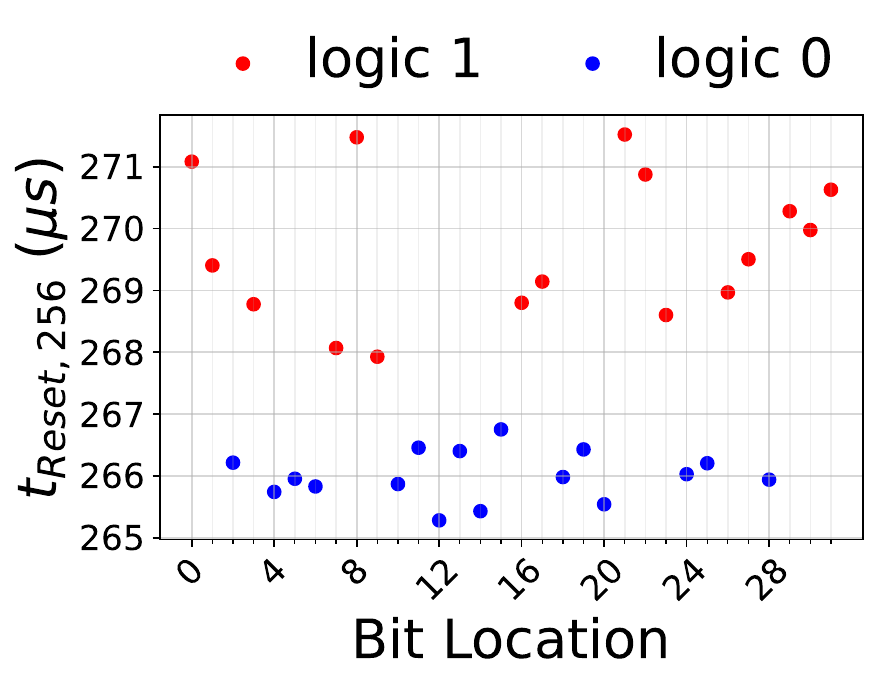}
        \caption{Over two months retention.}
        \label{fig:retention_2m_Reset}
    \end{subfigure}
    \caption{Retention characteristics of the hidden information using $t_{Reset,256}$.}
    \label{fig:retention_Reset}
\end{figure}

\subsection{Performance Analysis}

In this subsection, we use the encoding and retrieval time as well as encoding cost as the metric for measuring performance. 

\subsubsection{Encoding Time}

The proposed technique for information hiding relies on repeated
switching operation of ReRAM cells. Thus, the time required to encode the hidden information is directly proportional to the number of stress counts, $\mathcal{N}$. The estimated time to hide information is $\mathcal{T}_{encode} = (\mathcal{N} \times \mathcal{B}_{Msg} \times \mathcal{T}_{switch_{pair}})$, where $\mathcal{T}_{switch_{pair}} = (\mathcal{T}_{set} + \mathcal{T}_{reset})$ represents stressing time (\textit{set-reset} pair) for 256 addresses (switching resistance state with single \textit{write} command), and $\mathcal{B}_{Msg}$ represents the number of bits that need to conceal. The chip used for our experimental evaluation has the following timing parameters: $\mathcal{T}_{switch_{pair}} = (5ms +5ms) = 10ms$, and $\mathcal{B}_{Msg} = 32$. Thus, the baseline implementation requires $((5ms +5ms) \times 32 \times 15k) = 4800s$ for $15K$ switching operations to conceal secret message. Therefore, the throughput for the encoding process is $\frac{32bits}{4800s}= 0.4bit/min$. The hiding throughput will also be higher if $\mathcal{N}$ is smaller. Besides, it is worth mentioning that the encoding time of our proposed technique heavily depends on the \textit{write} speed of the ReRAM chips. Fortunately, in the past few years, the \textit{write} speed of ReRAM chips significantly improved and will continue to improve in the future. For example, the \textit{write} speed of \textit{MB85AS8MT} ReRAM chips is improved ${>}3X$ over its previous generation \textit{MB85AS4MT} ReRAM chips\footnote{\textit{MB85AS8MT} and \textit{MB85AS4MT} chips were launched in 2019 and 2016, respectively.}.

\subsubsection{Retrieval Time}
Unlike the encoding procedure, the extraction procedure is significantly faster. The estimated time to retrieve the hidden information can be calculated by\textemdash $\mathcal{T}_{retrieve} = (\mathcal{T}_{switch} \times \mathcal{B}_{Msg} \times \mathcal{N}_{rep})$, where $\mathcal{T}_{switch}$ is the average value of $t_{Set,256}$ or $t_{Reset,256}$, and $\mathcal{N}_{rep}$ represents the number of addresses used to encode single bits. After $15K$ stressing, the average value of $t_{Set,256}$ is ${\sim}250{\mu}s$, and we used $\mathcal{N}_{rep}=256$ in our implementation. Therefore, the throughput for the retrieval is $\frac{\mathcal{B}_{Msg}}{\mathcal{T}_{retrieve}}=\frac{32bits}{250{\mu}s \times 32 \times 256} = 15.625bits/s$. The average value $t_{Set,256}$ includes program data transfers from the microcontroller to the host computer and microcontroller overhead. Besides, the retrieval throughput will be higher if the hiding technique uses a smaller replica size (i.e., the number of memory addresses to encode each hidden bit), as discussed in Sect. \ref{subsec:grpSz}.

\subsubsection{Encoding Cost}
Our proposed technique requires a minimum of $15K$ \textit{set-reset} operations (i.e., $30K$ rewrite cycles) to make a distinguishable separation between logic `0' and `1' of the concealed information (using both $t_{Set,256}$ and $t_{Reset,256}$). However, the rated endurance of ReRAM chips is $1M$. Therefore, our proposed technique costs only $3\%$ of the rated endurance of encoded addresses.

\subsection{Initial Stress Tolerance}

The effectiveness of the proposed technique on moderately used memory chips is also examined. The influence of the initial stress count (i.e., the number of stress that occurred due to normal usage of memory before hiding information) is shown in Fig. \ref{fig:init_stress}. To simulate the normal usage of the storage device, we write random data patterns to the memory for the initial stressing. Random data patterns appear according to Ref. \cite{rand_op} to make more realistic stressing on the memory cells incurred from memory usage. For example, the separation between logic `0' and `1' at the initial stress level of $50K$ switching cycles shows the bit separation when bits are hidden using $15k$ \textit{set-reset} operation after performing $50K$ normal memory usage operations. We observe that bit separation decreases, i.e., bit error starts to occur (Fig. \ref{fig:init_stress_Reset}) as the initial stress level increases. However, increasing the stress level in the encoding process can tolerate a higher initial stress count. Besides, the \textit{set} operation can tolerate higher initial stress than the \textit{reset} operation. 
Note that the error rate observed using $t_{Reset,256}$ is still manageable ($3.125\%$) even after thousands of normal memory operations.

\begin{figure}[ht!]
    \centering
    \captionsetup{justification=centering, margin= 0.5cm}
    \begin{subfigure}[t]{0.227\textwidth}
        \centering
        \includegraphics[trim=0.2cm 0.3cm 0.2cm 0.5cm, clip, width = 0.9\textwidth]{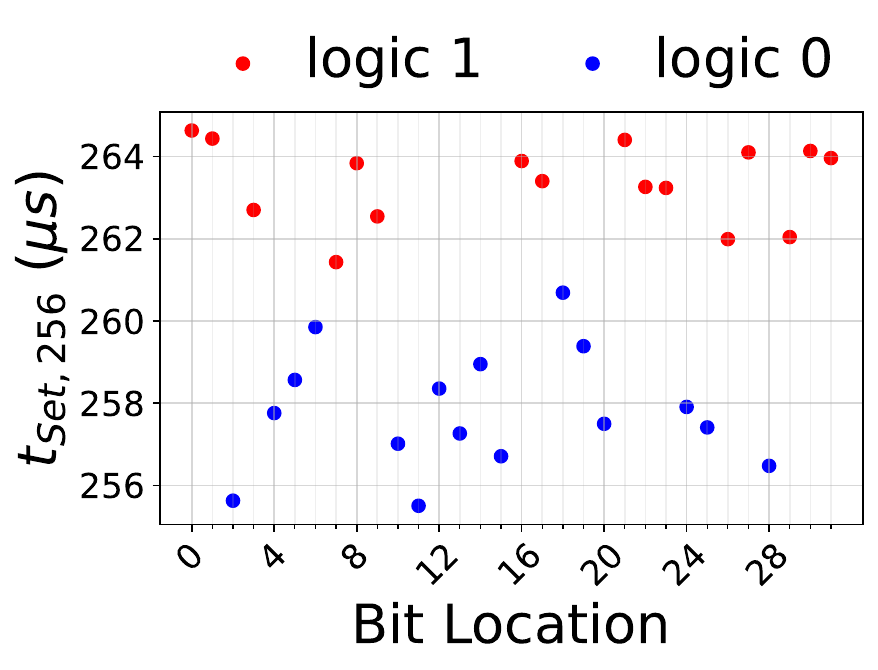}
        \caption{$\mathcal{N}=15K$}
        \label{fig:init_stress_Set}
    \end{subfigure}
    \begin{subfigure}[t]{0.235\textwidth}
        \centering
        \includegraphics[trim=0.2cm 0.3cm 0.2cm 0.5cm, clip, width = 0.9\textwidth]{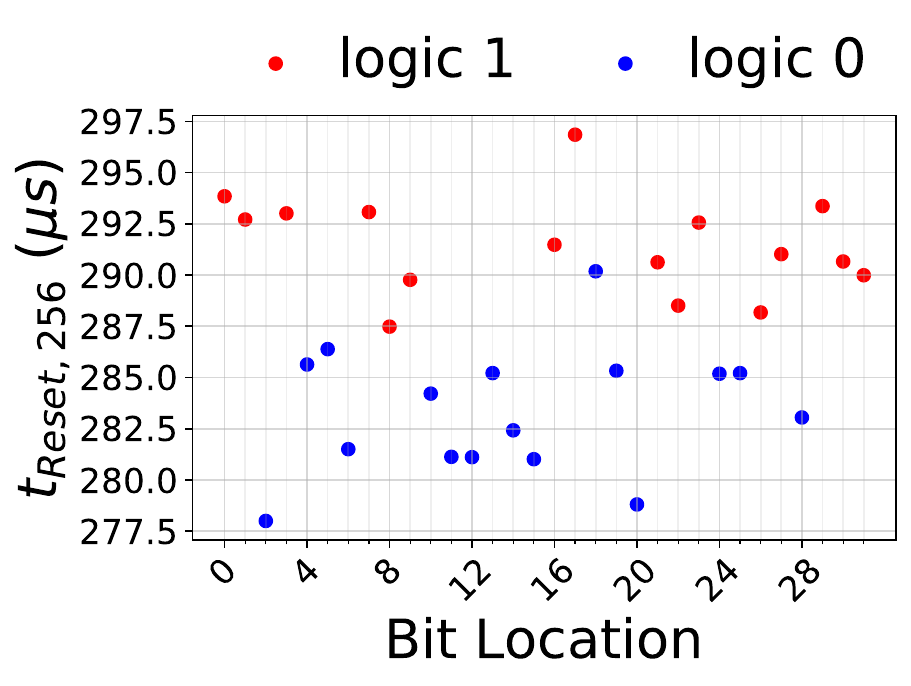}
        \caption{$\mathcal{N}=15K$}
        \label{fig:init_stress_Reset}
    \end{subfigure}
    \caption{Influence of initial stress ($50K$) on the hidden information using- (a) $t_{Set,256}$ and (b) $t_{Reset,256}$.}
    \label{fig:init_stress}
\end{figure}

\subsection{Post-Hiding Stress Tolerance} \label{subsec:post_stress}

To test the post-hiding stress tolerance of our proposed technique, we deliberately stress the memory chip after hiding information on the chip. We write random data patterns to the memory to emulate the post-hiding stressing. The occurrence of random data patterns appears according to Ref. \cite{rand_op} to make more realistic stressing on the memory cells incurred from memory usage. Fig. \ref{fig:T_postStress} and Fig. \ref{fig:stress} illustrate the influence of post-hiding stressing (i.e., the number of stresses performed after hiding information). 

Fig. \ref{fig:T_postStress} represents the post-hiding stress tolerance of the hidden data where a minimum ($\mathcal{N} = 15K$) stress count is used to hide the information\footnote{Also verified for other stressing levels (up to $\mathcal{N} = 45K$) used to hide information.}. We conceal the data in a confidential memory location using the key. The red and blue dots represent the hidden logic 1s and 0s, respectively. Fig. \ref{fig:T_postStress} shows that logic `1' and logic `0' remain separated up to $130K$ stress count (Fig. \ref{fig:SetT_postStress_130K}). They become inseparable at $140K$ stress count (Fig. \ref{fig:SetT_postStress_140K}). Similarly, with $t_{Reset,256}$, logic `1' and logic `0' remain separated up to $40K$ stress count (Fig. \ref{fig:ResetT_postStress_40K}) and become inseparable at $50K$ stress count (Fig. \ref{fig:ResetT_postStress_50K}). 

\begin{figure}[ht!]
\centering
\captionsetup{justification=centering, margin= 0cm}
\begin{minipage} []{.24\textwidth}
    \centering
    \captionsetup{justification=centering, margin= 0cm}
    \begin{subfigure}[t]{1\textwidth}
        \centering
        \includegraphics[trim=0cm 0cm 0cm 0cm, clip, width=0.9\textwidth]{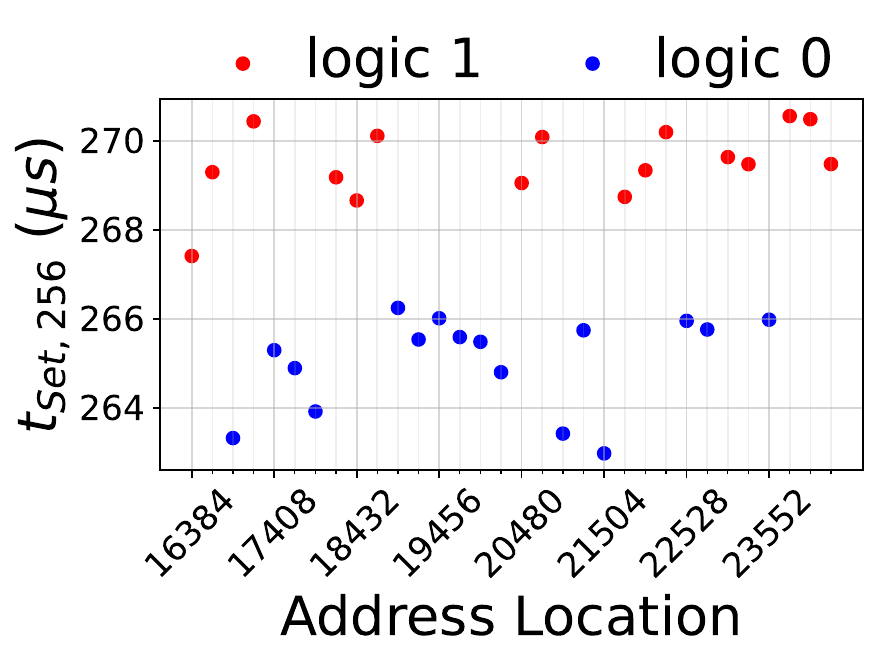}
        \caption{Post Stressing = $130K$}
        \label{fig:SetT_postStress_130K}
    \end{subfigure}%
    \vspace{\medskipamount}
    \begin{subfigure}[t]{1\textwidth}
        \centering
        \includegraphics[trim=0cm 0cm 0cm 0cm, clip, width = 0.9\textwidth]{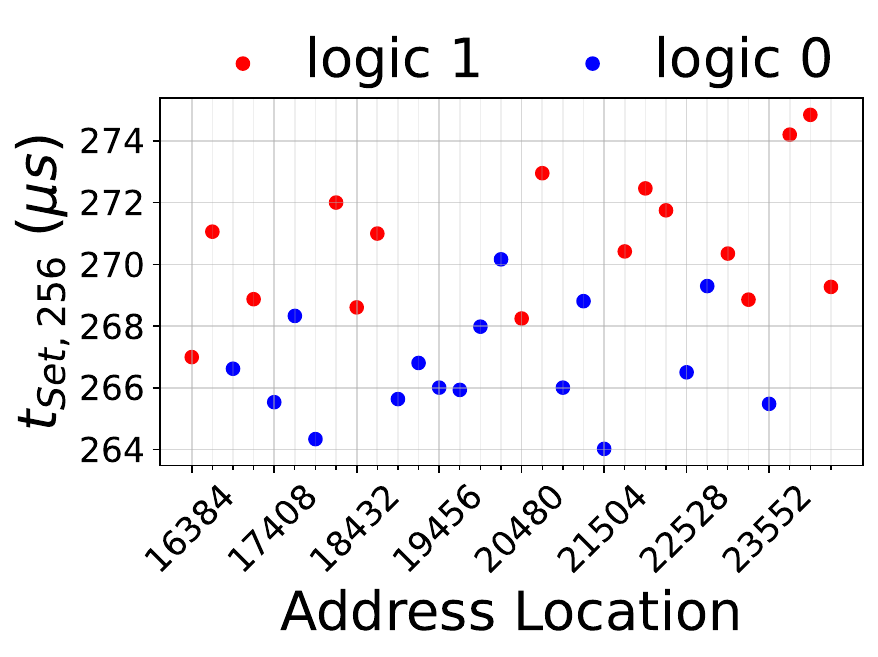}
        \caption{Post Stressing = $140K$}
        \label{fig:SetT_postStress_140K}
    \end{subfigure}
\end{minipage}
\begin{minipage} []{.24\textwidth}
    \centering
    \captionsetup{justification=centering, margin= 0cm}
    \begin{subfigure}[t]{1\textwidth}
        \centering
        \includegraphics[trim=0cm 0cm 0cm 0cm, clip, width=0.9\textwidth]{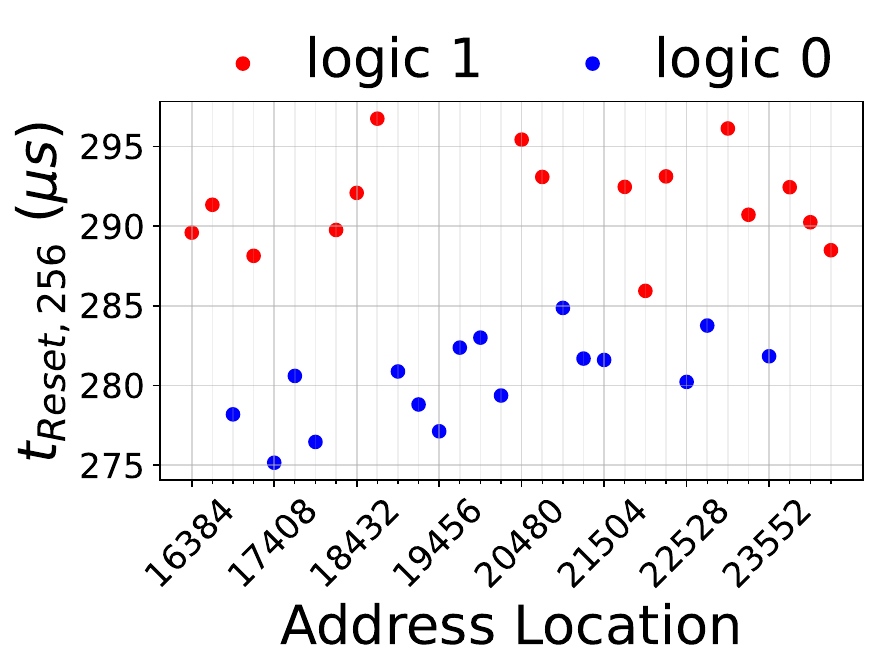}
        \caption{Post Stressing = $40K$}
        \label{fig:ResetT_postStress_40K}
    \end{subfigure}%
    \vspace{\medskipamount}
    \begin{subfigure}[t]{1\textwidth}
        \centering
        \includegraphics[trim=0cm 0cm 0cm 0cm, clip, width = 0.9\textwidth]{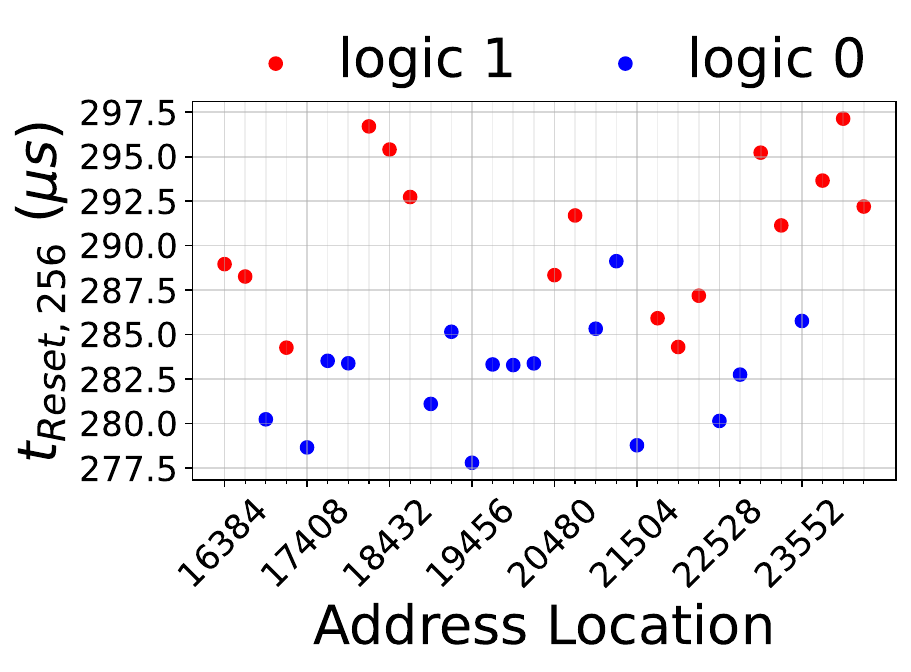}
        \caption{Post Stressing = $50K$}
        \label{fig:ResetT_postStress_50K}
    \end{subfigure}
\end{minipage}
\caption{Hidden data (hiding stress count, $\mathcal{N} = 15K$) at different post-hiding stress level- (a)--(b) $t_{Set,256}$ at stress count $130K$, and $140K$; (c)--(d) $t_{Reset,256}$ at stress count $40K$, and $50K$.}
\label{fig:T_postStress}
\end{figure}

In addition, Fig. \ref{fig:stress} represents the distribution of $d(b_0,b_1)$ at different post-hiding stress levels, where $d(b_0,b_1)$ represents the distance between logic `0' bits ($b_0$) and logic `1' bits ($b_1$). Each dot in Fig. \ref{fig:stress} represents $d(b_0^i,b_1^j)$ for each possible $(b_0^i,b_1^j)$. The distance should be positive for well-separated logic `0' and `1'. A larger value of $d(b_0^i,b_1^j)$ is more desirable as it provides better separation between logic `0' and logic `1' bits. However, if the maximum value of \textit{set/reset} time of logic `0' bits is larger than the minimum value of \textit{set/reset} time of logic `1' bits (similar to Fig. \ref{fig:SetT_postStress_140K}), then logic `0' bits and logic `1' bits cannot be separated with 100\% accuracy. In such a scenario, the $d(b_0^i,b_1^j)$ can be negative for a few pairs of $(b_0^i,b_1^j)$. The figure demonstrates that the separation between logic `0' bits and logic `1' bits degrade with respect to post-stress count. However, the separation between logic `0' bits and logic `1' bits is quite reasonable, even after thousands of post-hiding stresses. For example, the logic `0' bits ($b_0$) and logic `1' bits ($b_1$) are clearly separable up to $110K$, $140K$, and $230K$, post-hiding stress levels using $t_{Set,256}$ with $\mathcal{N} = 15K$ (Fig. \ref{fig:SetT_stress_15K}), $30K$ (Fig. \ref{fig:SetT_stress_30K}), and $45K$ (Fig. \ref{fig:SetT_stress_45K}) hiding stress count used to conceal information, respectively (i.e., $\min \big( d(b_0^i,b_1^j) \big) > 0$). On the other hand, the logic `0' bits ($b_0$) and logic `1' bits ($b_1$) are clearly separable up to $40K$, $70K$, and $140K$, post-hiding stress levels using $t_{Reset,256}$ with $\mathcal{N} = 15K$ (Fig. \ref{fig:ResetT_stress_15K}), $30K$ (Fig. \ref{fig:ResetT_stress_30K}), and $45K$ (Fig. \ref{fig:ResetT_stress_45K}) hiding stress count, respectively. However, note that if we tolerate one or two-bit errors, then our proposed scheme can survive a few more thousands of post-hiding stress levels (Table \ref{Tab:summ1}).

\begin{figure}[ht!]
\centering
\captionsetup{justification=centering, margin= 0cm}
\begin{minipage} []{.235\textwidth}
    \centering
    \captionsetup{justification=centering, margin= 0cm}
    \begin{subfigure}[t]{1\textwidth}
        \centering
        \includegraphics[trim=0.1cm 0.1cm 0.1cm 0.1cm, clip, width=0.9\textwidth]{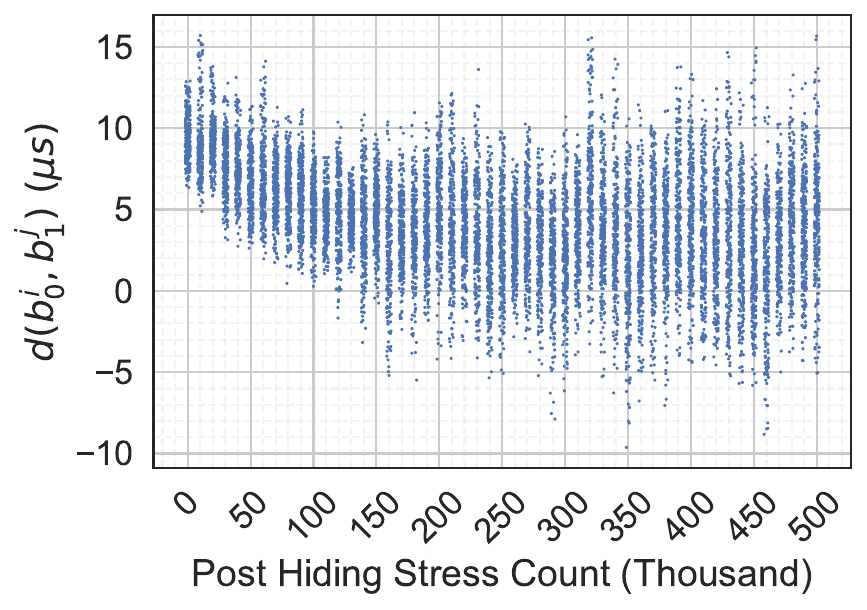}
        \caption{$\mathcal{N}=15K$}
        \label{fig:SetT_stress_15K}
    \end{subfigure}%
    \vspace{\medskipamount}
    \begin{subfigure}[t]{1\textwidth}
        \centering
        \includegraphics[trim=0.1cm 0.1cm 0.1cm 0.1cm, clip, width = 0.9\textwidth]{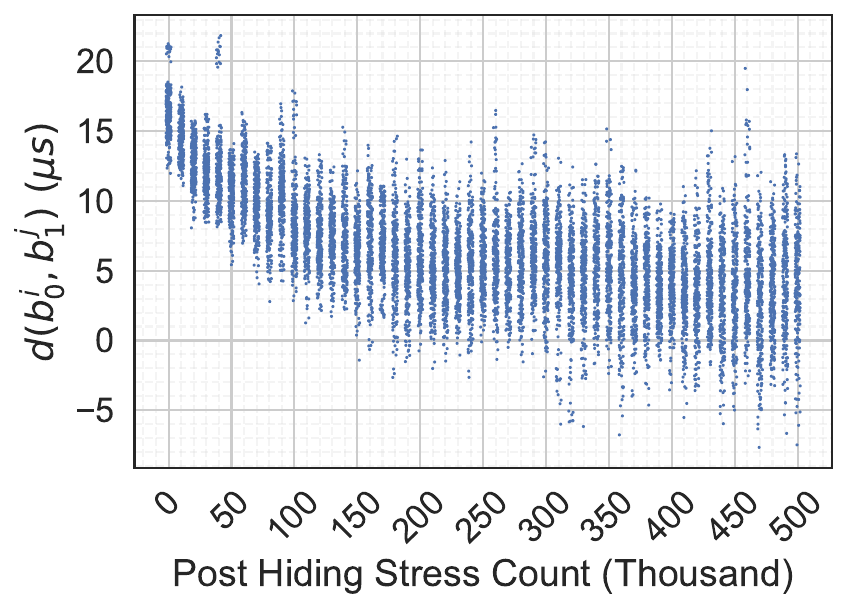}
        \caption{$\mathcal{N}=30K$}
        \label{fig:SetT_stress_30K}
    \end{subfigure}%
    \vspace{\medskipamount}
    \begin{subfigure}[t]{1\textwidth}
        \centering
        \includegraphics[trim=0.1cm 0.1cm 0.1cm 0.1cm, clip, width = 0.9\textwidth]{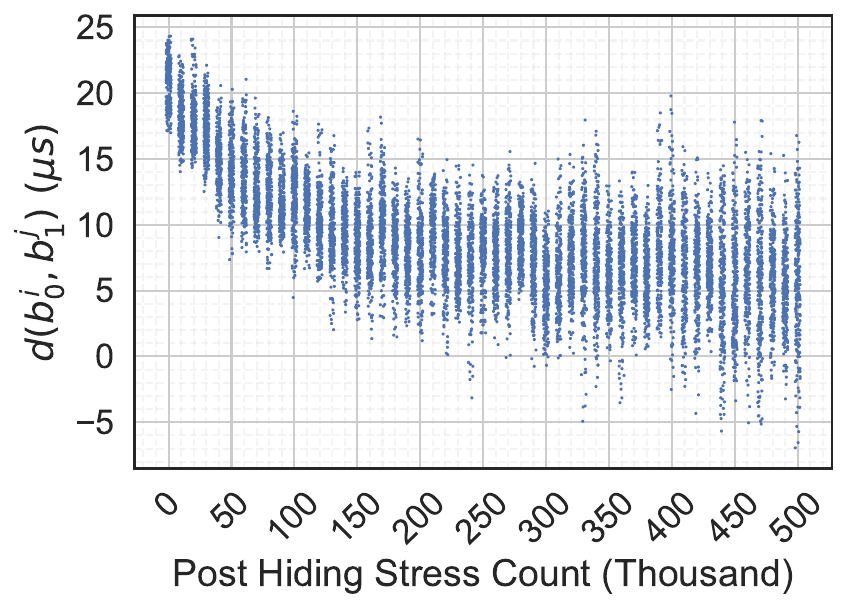}
        \caption{$\mathcal{N}=45K$}
        \label{fig:SetT_stress_45K}
    \end{subfigure}
\end{minipage}
\begin{minipage} []{.235\textwidth}
    \centering
    \captionsetup{justification=centering, margin= 0cm}
    \begin{subfigure}[t]{1\textwidth}
        \centering
        \includegraphics[trim=0.1cm 0.1cm 0.1cm 0.1cm, clip, width=0.9\textwidth]{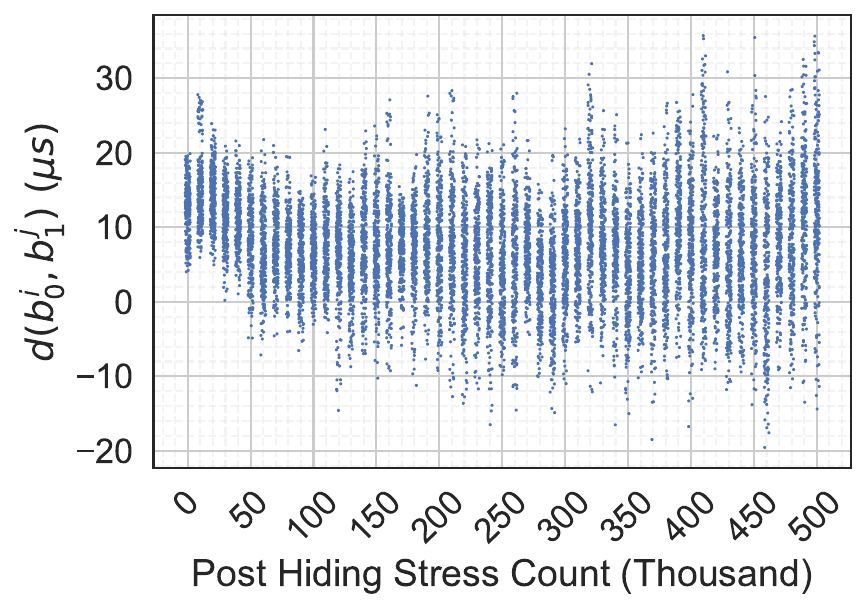}
        \caption{$\mathcal{N}=15K$}
        \label{fig:ResetT_stress_15K}
    \end{subfigure}%
    \vspace{\medskipamount}
    \begin{subfigure}[t]{1\textwidth}
        \centering
        \includegraphics[trim=0.1cm 0.1cm 0.1cm 0.1cm, clip, width = 0.9\textwidth]{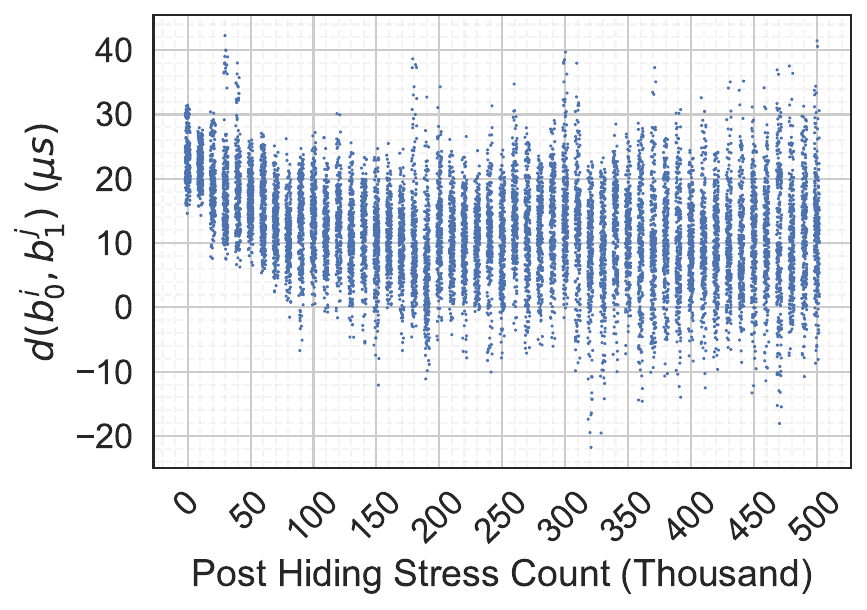}
        \caption{$\mathcal{N}=30K$}
        \label{fig:ResetT_stress_30K}
    \end{subfigure}%
    \vspace{\medskipamount}
    \begin{subfigure}[t]{1\textwidth}
        \centering
        \includegraphics[trim=0.1cm 0.1cm 0.1cm 0.1cm, clip, width = 0.9\textwidth]{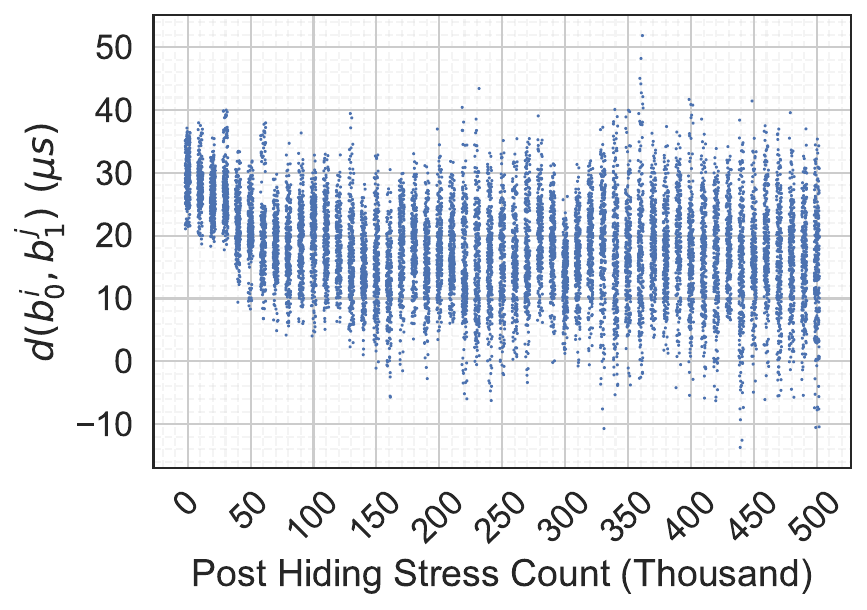}
        \caption{$\mathcal{N}=45K$}
        \label{fig:ResetT_stress_45K}
    \end{subfigure}
\end{minipage}
\caption{Post-hiding stress tolerance of concealed data at different hiding stress levels, using-\\ (a)--(c) $t_{Set,256}$, and (d)--(f) $t_{Reset,256}$.}
\label{fig:stress}
\end{figure}

Table \ref{Tab:summ} summarizes the post-hiding stress tolerance statistics for all test chips. The first two columns show the stress level used to hide information and the switching (\textit{set} or \textit{reset}) operation considered while performing post-hiding stressing. Finally, column three to seven represents the post-hiding stress levels that different chips can endure. We observe that more stress in the hiding process increases the write time difference between bits hiding `1's and `0's. If we use lower stress levels (even lower than $15K$) to encode information, the hidden information can survive fewer thousands of normal memory usage operations. Besides, the \textit{set} operation can endure a higher post-hiding stress count than the \textit{reset} operation. Therefore, depending on the application requirement, one can choose the stressing level to encode information to sustain the desired level of memory usage. \textcolor{black}{Note that the \textit{set} operation can tolerate both higher initial stress and higher post-hiding stress. In ReRAM, the \textit{set} operation means transitioning from HRS to LRS, whereas the \textit{reset} operation means transitioning from LRS to HRS (Fig. \ref{fig:ReRAM}). Repeated stressing on ReRAM gradually reduces the oxide layer integrity and introduces crystal defects (dielectric breakdown). Due to this crystal defect, the oxide layer contains more charge carriers than when it was new. Therefore, the resistance of the HRS reduces over time if we keep the same operating condition as the fresh ReRAM cell. Additionally, the oxide layer of ReRAM represents its highest resistance when there is no crystal defect. As the crystal defect introduced by device usage is random, the \textit{reset} operation becomes noisier over time compared to the \textit{set} operation ($2\times$ noisier \cite{feng2016investigation}). Therefore, the \textit{reset} operation tolerates less initial stress and less post-hiding stress.}

\begin{table}[ht!]
\caption{Post-hiding stress tolerance (considering zero-bit error).}
\setcellgapes{1pt}
\captionsetup{justification=centering, margin= 0cm}
\makegapedcells
\centering
\setlength\tabcolsep{5pt} 
\resizebox{0.48\textwidth}{!}
{
    \begin{tabular}{|c|l|c|c|c|c|c|}
    \hline
    \multicolumn{1}{|l|}{\begin{tabular}[c]{@{}l@{}}Stress \\ Count, $\mathcal{N}$\end{tabular}} & Op. & Chip1 & Chip2 & Chip3 & Chip4 & Chip5 \\ \hline
    \multirow{2}{*}{15K} & Set   & 110K & 130K & 100K & 100K & 130K \\ \cline{2-7} 
                         & Reset & 60K  & 40K  & 50K  & 30K  & 40K  \\ \hline
    \multirow{2}{*}{30K} & Set   & 150K & 150K & 140K & 150K & 140K \\ \cline{2-7} 
                         & Reset & 70K  & 100K & 100K & 90K  & 80K  \\ \hline
    \multirow{2}{*}{45K} & Set   & 180K & 190K & 190K & 200K & 230K \\ \cline{2-7} 
                         & Reset & 180K & 180K & 200K & 130K & 140K \\ \hline
    \end{tabular}
}
\label{Tab:summ}
\end{table}

How long the confidential information needs to be concealed entirely relies on the application. If we use higher stress levels to hide the data, those hidden messages can tolerate more post-hiding stress. For example, if we want to keep the hidden message up to the full endurance level of the memory, we need to stress a little higher than $45K$ (which we demonstrated in Fig \ref{fig:SetT_stress_45K}). If we hide the date with $45K$ stressing, it can tolerate $180K$ post-hiding stressing ($36\%$ of the rated endurance of ReRAM chips) without ECC (Table \ref{Tab:summ}) and $430K$ post-hiding stressing ($86\%$\footnote{As with higher stressing levels and without ECC, the hidden message remains concealed for $36\%$ of the rated endurance of ReRAM chips; Hence we avoided such complex and extensive ECC circuits in this work. However, we noted the reviewer’s comment for our future research.} of the rated endurance) with ECC (Table \ref{Tab:summ1}), respectively. In addition, if the secret message needs not to be kept for a higher endurance level, we can consider a lesser stressing level, such as $15K$, to hide the data to make the secret message unrecoverable after $20\%$ (Table \ref{Tab:summ}) of the rated endurance of ReRAM chips. Furthermore, if we want to keep the hidden data longer with fewer stressing counts, we can continue stressing those cells before the hidden data disappears.

\begin{table}[ht!]
\caption{Worst-case post-hiding stress tolerance (considering \textit{set} operation, $t_{Set,256}$).}
\setcellgapes{1pt}
\captionsetup{justification=centering, margin= 0cm}
\makegapedcells
\centering
\setlength\tabcolsep{5pt} 
\resizebox{0.3\textwidth}{!}
{
    \begin{tabular}{|c|ccc|}
    \hline
    \multirow{2}{*}{\begin{tabular}[c]{@{}c@{}}Stress \\ Count, $\mathcal{N}$\end{tabular}} & \multicolumn{3}{c|}{Bit Error Count} \\ \cline{2-4} 
        & \multicolumn{1}{c|}{0}    & \multicolumn{1}{c|}{1}    & 2    \\ \hline
    15k & \multicolumn{1}{c|}{100K} & \multicolumn{1}{c|}{150K} & 200K \\ \hline
    30k & \multicolumn{1}{c|}{140K} & \multicolumn{1}{c|}{250K} & 310K \\ \hline
    45k & \multicolumn{1}{c|}{180K} & \multicolumn{1}{c|}{400K} & 430K \\ \hline
    \end{tabular}
}
\label{Tab:summ1}
\end{table}

Besides, Table \ref{Tab:summ1} represents the minimum error tolerance of our proposed data-hiding scheme. The \nth{1} column represents the number of stresses used to hide the data (considering \textit{set} operation, $t_{Set,256}$). The \nth{2} – \nth{4} columns represent the minimum number of allowable post-hiding stresses considering $0-, 1-$, and $2-$bit errors, respectively. Table \ref{Tab:summ1} shows that if $15K$ stress level is used to hide the data, those secret messages can tolerate $100K$ ($20\%$ of the rated endurance), $150K$ ($30\%$ of the rated endurance), and $200K$ ($40\%$ of the rated endurance) post-hiding stressing with $0-, 1-$, and $2-$bit error(s), respectively. On the other hand, if $45K$ stress level is used to hide the data considering \textit{set} operation ($t_{Set,256}$), those secret messages can tolerate $180K$ ($36\%$ of the rated endurance), $400K$ ($80\%$ of the rated endurance), and $430K$ ($86\%$ of the rated endurance) post-hiding stressing with $0-, 1-$, and $2-$bit error(s), respectively.

\begin{figure}[ht!]
    \centering
    \captionsetup{justification=centering, margin= 0.5cm}
    \begin{subfigure}[t]{0.227\textwidth}
        \centering
        \includegraphics[trim=0.1cm 0.1cm 0.1cm 0.1cm, clip, width = 0.9\textwidth]{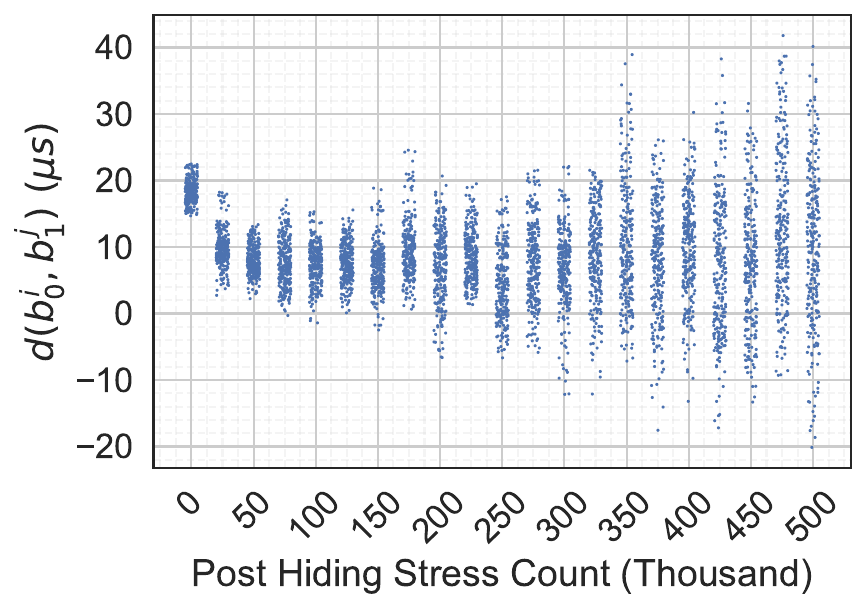}
        \caption{$\mathcal{N}=45K$}
        \label{fig:post_stress_10_Set}
    \end{subfigure}
    \begin{subfigure}[t]{0.235\textwidth}
        \centering
        \includegraphics[trim=0.1cm 0.1cm 0.1cm 0.1cm, clip, width = 0.9\textwidth]{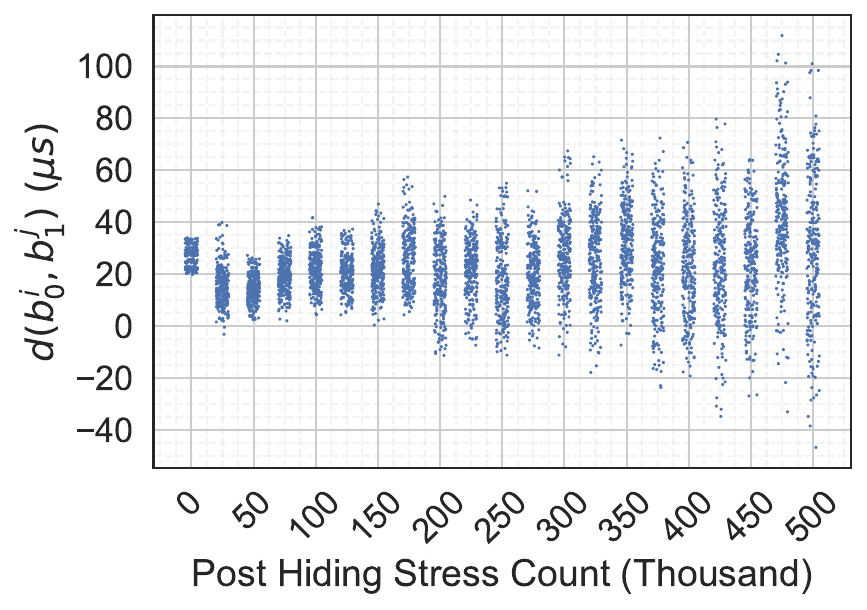}
        \caption{$\mathcal{N}=45K$}
        \label{fig:post_stress_10_Reset}
    \end{subfigure}
    \caption{Post-hiding stress tolerance of concealed data considering maximum stressing using- (a) $t_{Set,256}$ and (b) $t_{Reset,256}$.}
    \label{fig:post_stress_10}
\end{figure}

Furthermore, to apply maximum post-hiding stress, we write `1' $\rightarrow$ `0' $\rightarrow$ `1' to each memory cell after hiding information on the memory chip. The influence of maximum post-hiding stress on the separation between logic bits vs. the number of post-hiding stresses performed after hiding information is shown in Fig. \ref{fig:post_stress_10}. From the figure, we can see that the distance between logic `1' and `0' decreases as the post-hiding stress level increases. However, the distance between logic `1' and `0' of hidden information is quite reasonable for both $t_{Set,256}$ and $t_{Reset,256}$, even after thousands of post-hiding stress counts. For example, with $45K$ switching operations used to hide information can endure $200K$ post-hiding stress levels for both $t_{Set,256}$ (Fig. \ref{fig:post_stress_10_Set}) and $t_{Reset,256}$ (Fig. \ref{fig:post_stress_10_Reset}).

\textcolor{black}{Note that, as the \textit{set/reset} time increases monotonically with respect to usage (Fig. \ref{fig:char}), the \textit{set/reset} time of the imprinted cells should also shift upward with regular usage. In this scenario, the bit `0' and bit `1' can be differentiated by either of the following two techniques-}
\begin{itemize}[leftmargin=*, topsep=0pt,itemsep=-1ex,partopsep=1ex,parsep=1ex]
    \item \textcolor{black}{\textbf{Using clustering algorithm:} As we determine the bits `0' and `1' based on the distance in \textit{set/reset} time, a simple clustering algorithm such as k-mean clustering is extremely effective in clustering all `0' bits and `1' bits. Therefore, we do not need to specify any threshold value explicitly.}
    \item \textcolor{black}{\textbf{Threshold shifting:} If the clustering algorithm is not available due to the device limitation, the user can monitor the \textit{set/reset} time of a set of memory cells (reference cells) where the hidden data in not imprinted. Now, based on the \textit{set/reset} time on the reference cell, one can easily determine the current usage level of the ReRAM chips and shift the threshold accordingly. Ideally, the \textit{set/reset} time of these reference cells should be in a similar range as of the logic `0' imprinted memory cells.}
\end{itemize}

\textcolor{black}{Lastly, In our experiment, we have emulated the worst-case scenario of memory stress (i.e., toggling memory bit in the write cycle). However, in reality, a memory bit does not toggle on every write operation. For example, if we overwrite a random byte, `00101000', with another random byte, `00110101', only four bits will be toggled, and others will not experience any stress. Therefore, we expect our hidden data to last for a much higher number of write cycles than we demonstrated in our above discussion. Additionally, in a practical scenario, memory cell usually experiences a specific usage pattern that has been extensively explored in previous work (\cite{rand_op}). Under regular usage, the most significant bits (MSB) usually experience less stress than the least significant bits (LSB), whereas the usage pattern over the entire memory address follows almost a uniform distribution. In our proposed technique, we measure the \textit{set/reset} time for the whole byte; therefore, the usage deviation from MSB to LSB should not impact our proposed technique.}
\subsection{Robustness Analysis}

The hidden message should be resilient to the variation of operating conditions, i.e., it will not be possible to modify or change the hidden information with localized heating or operating voltage. Inherently, all modern ICs are resilient to small variations in operating voltage as they are usually integrated with a voltage regulator. Voltage regulators can retain the operating voltage within a valid range of supply voltage. However, to verify the robustness of our encoding technique against the temperature with post-hiding stressing, first, we conceal information in a confidential address space with $15K, ~30K$, and $45K$ stress levels, respectively. Then, we write random data patterns to the memory for $50K$ to $200K$ times with $30K$ intervals to examine the robustness of the proposed hiding scheme. Random data patterns appear according to Ref. \cite{rand_op} to make more realistic stressing on the memory cells incurred from memory usage. After every $30K$ memory operation, we isolated the memory chip from the system and baked it at $80^{\circ}C$ for $1$ day. Lastly, we have evaluated the $t_{Set,256}$ and $t_{Reset,256}$ while maintaining the chip temperature of $80^{\circ}C$ for all three stress levels (in our case, $\mathcal{N} = 15K,~30K$, and $45K$, respectively) used to conceal information.

\begin{figure}[ht!]
\centering
\captionsetup{justification=centering, margin= 0cm}
\begin{minipage} []{.235\textwidth}
    \centering
    \captionsetup{justification=centering, margin= 0cm}
    \begin{subfigure}[t]{1\textwidth}
        \centering
        \includegraphics[trim=0.1cm 0.1cm 0.1cm 0.1cm, clip, width=0.9\textwidth]{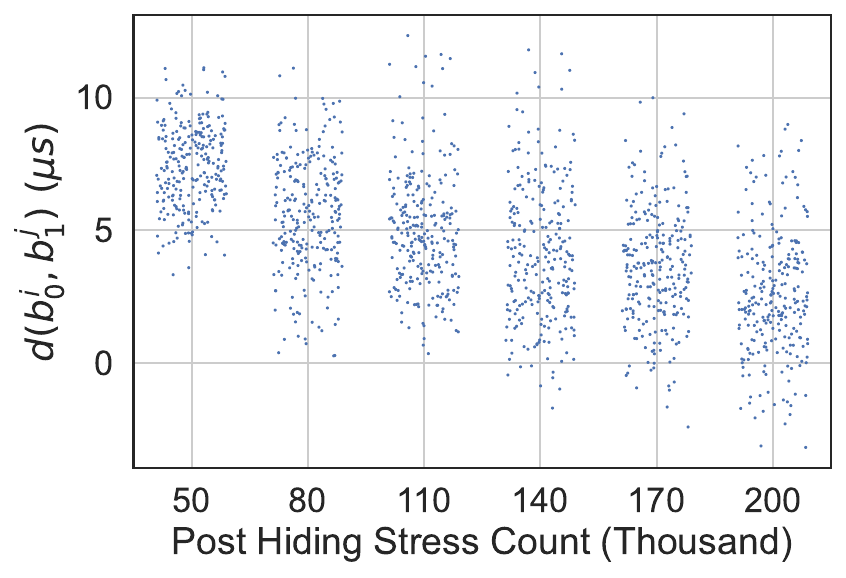}
        \caption{$\mathcal{N}=15K$}
        \label{fig:SetT_robust_15K}
    \end{subfigure}%
    \vspace{\medskipamount}
    \begin{subfigure}[t]{1\textwidth}
        \centering
        \includegraphics[trim=0.1cm 0.1cm 0.1cm 0.1cm, clip, width = 0.9\textwidth]{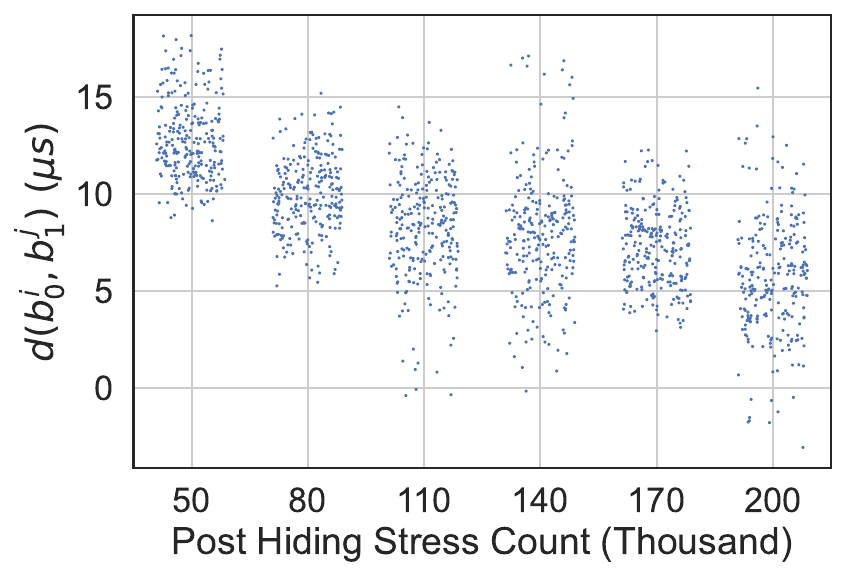}
        \caption{$\mathcal{N}=30K$}
        \label{fig:SetT_robust_30K}
    \end{subfigure}%
    \vspace{\medskipamount}
    \begin{subfigure}[t]{1\textwidth}
        \centering
        \includegraphics[trim=0.1cm 0.1cm 0.1cm 0.1cm, clip, width = 0.9\textwidth]{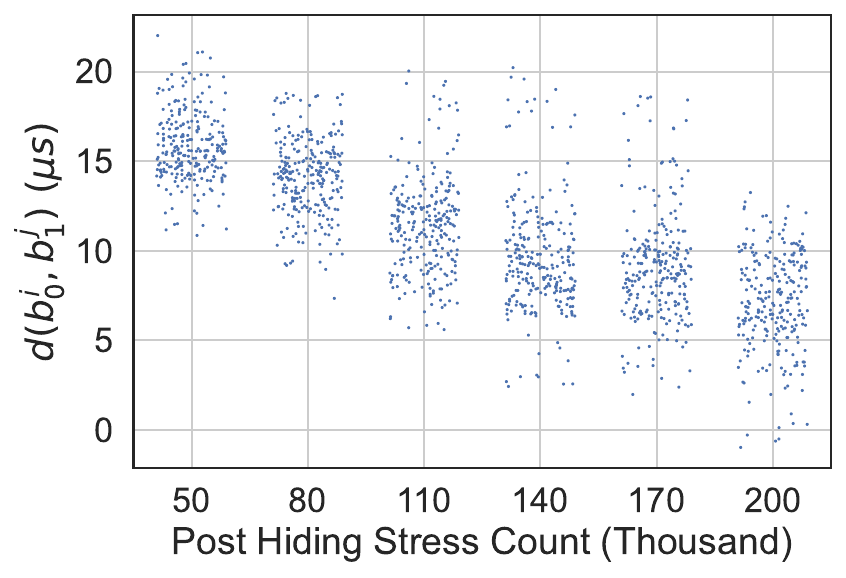}
        \caption{$\mathcal{N}=45K$}
        \label{fig:SetT_robust_45K}
    \end{subfigure}
\end{minipage}
\begin{minipage} []{.235\textwidth}
    \centering
    \captionsetup{justification=centering, margin= 0cm}
    \begin{subfigure}[t]{1\textwidth}
        \centering
        \includegraphics[trim=0.1cm 0.1cm 0.1cm 0.1cm, clip, width=0.9\textwidth]{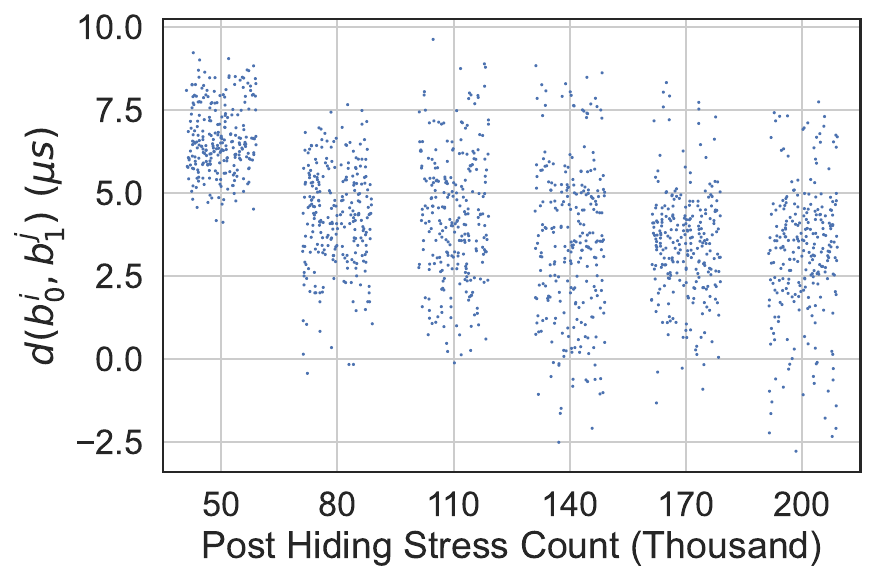}
        \caption{$\mathcal{N}=15K$}
        \label{fig:SetT_robust_15K_HT}
    \end{subfigure}%
    \vspace{\medskipamount}
    \begin{subfigure}[t]{1\textwidth}
        \centering
        \includegraphics[trim=0.1cm 0.1cm 0.1cm 0.1cm, clip, width = 0.9\textwidth]{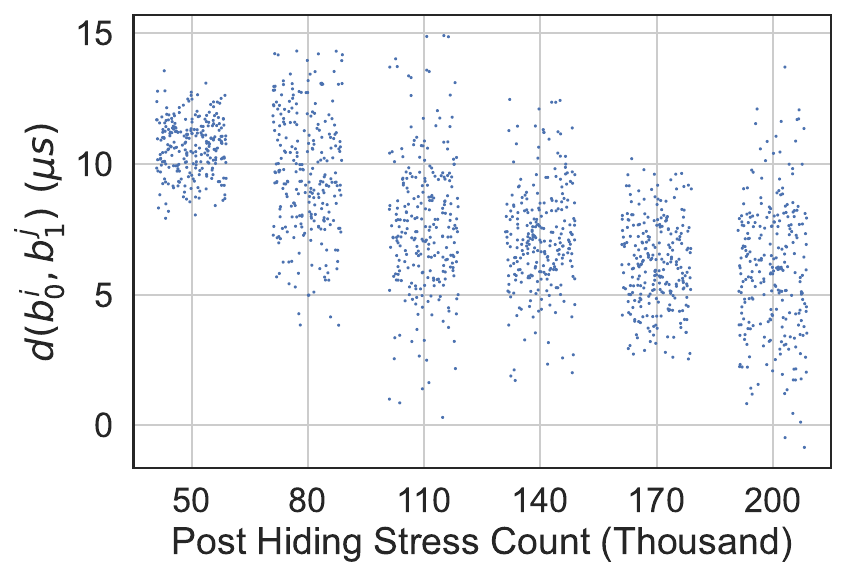}
        \caption{$\mathcal{N}=30K$}
        \label{fig:SetT_robust_30K_HT}
    \end{subfigure}%
    \vspace{\medskipamount}
    \begin{subfigure}[t]{1\textwidth}
        \centering
        \includegraphics[trim=0.1cm 0.1cm 0.1cm 0.1cm, clip, width = 0.9\textwidth]{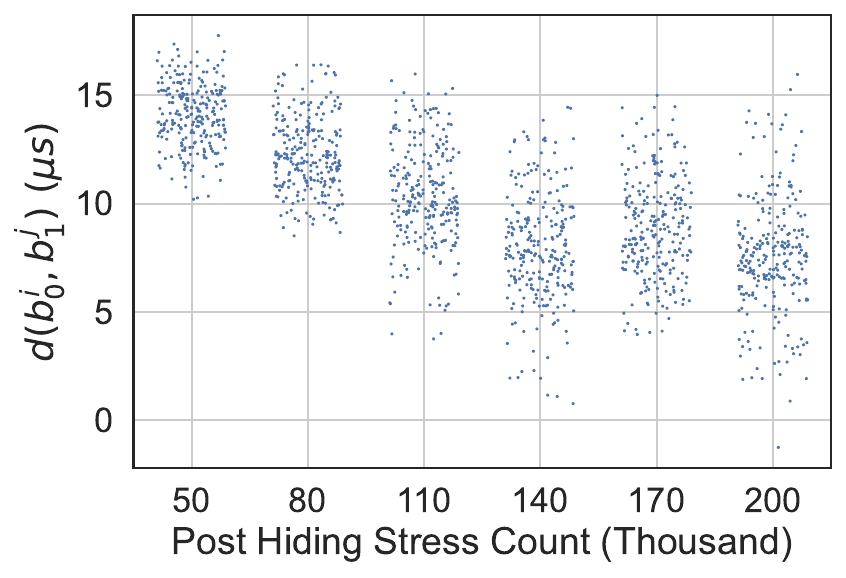}
        \caption{$\mathcal{N}=45K$}
        \label{fig:SetT_robust_45K_HT}
    \end{subfigure}
\end{minipage}
\caption{Robustness analysis (a)--(c) before (d)--(f) after high-temperature baking ($80^{\circ}C$) with- $t_{Set,256}$.}
\label{fig:robust_set}
\end{figure}

Fig. \ref{fig:robust_set} and Fig. \ref{fig:robust_reset} illustrate the influence of high-temperature baking. These figures represent the distribution of $d(b_0,b_1)$ at different post-hiding stress levels, where $d(b_0,b_1)$ represents the distance between logic `0' bits ($b_0$) and logic `1' bits ($b_1$). As discussed in Sect. \ref{subsec:post_stress}, a larger positive value of $d(b_0^i,b_1^j)$ is desirable to better separate logic `0' and logic `1' bits. However, in Fig. \ref{fig:ResetT_robust_15K} and Fig. \ref{fig:ResetT_robust_15K_HT}, the $d(b_0^i,b_1^j)$ is negative for a few pairs of $(b_0^i,b_1^j)$; hence, logic `0' bits and logic `1' bits cannot be separated with 100\% accuracy. Here, Figs. \ref{fig:SetT_robust_15K}, \ref{fig:SetT_robust_30K}, and \ref{fig:SetT_robust_45K} show $d(b_0^i,b_1^j)$ using $t_{Set,256}$ before performing high-temperature system-level operation. Similarly, Figs. \ref{fig:SetT_robust_15K_HT}, \ref{fig:SetT_robust_30K_HT}, and \ref{fig:SetT_robust_45K_HT} show $d(b_0^i,b_1^j)$ using $t_{Set,256}$ after performing high-temperature baking and high-temperature system-level operation. Similar observations are valid for
$t_{Reset,256}$ (Fig. \ref{fig:robust_reset}). We have observed that the hidden information is not significantly affected by temperature and remains well-separated after the high-temperature baking and high-temperature system-level operation (considering both $t_{Set,256}$ and $t_{Reset,256}$). Such behavior of ReRAM is expected as the resistance ratio ($\frac{R_{HRS}}{R_{LRS}}$) is relatively temperature insensitive \cite{Bogdan:ReRAM}. Note that the ReRAM chips used in our experiment are rated to operate up to $85^{\circ}C$.

\begin{figure}[ht!]
\centering
\captionsetup{justification=centering, margin= 0cm}
\begin{minipage} []{.235\textwidth}
    \centering
    \captionsetup{justification=centering, margin= 0cm}
    \begin{subfigure}[t]{1\textwidth}
        \centering
        \includegraphics[trim=0.1cm 0.1cm 0.1cm 0.1cm, clip, width=0.9\textwidth]{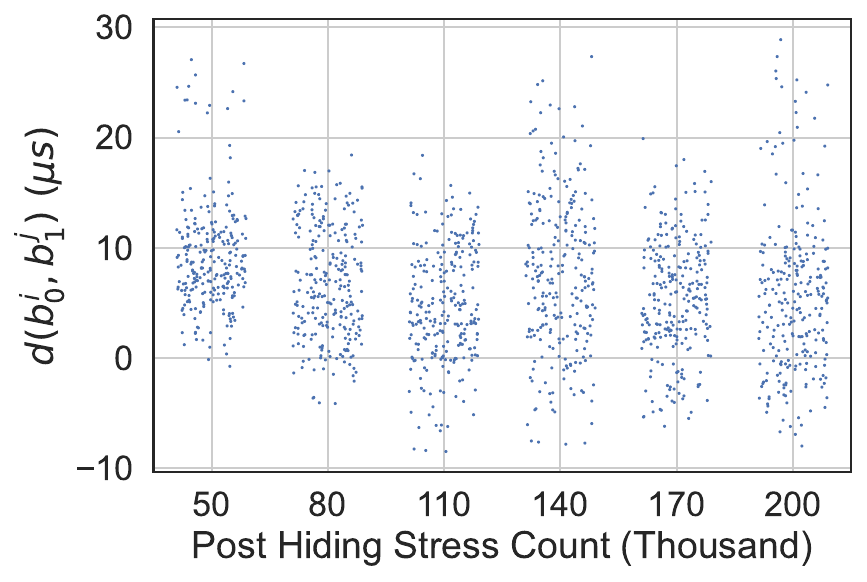}
        \caption{$\mathcal{N}=15K$}
        \label{fig:ResetT_robust_15K}
    \end{subfigure}%
    \vspace{\medskipamount}
    \begin{subfigure}[t]{1\textwidth}
        \centering
        \includegraphics[trim=0.1cm 0.1cm 0.1cm 0.1cm, clip, width = 0.9\textwidth]{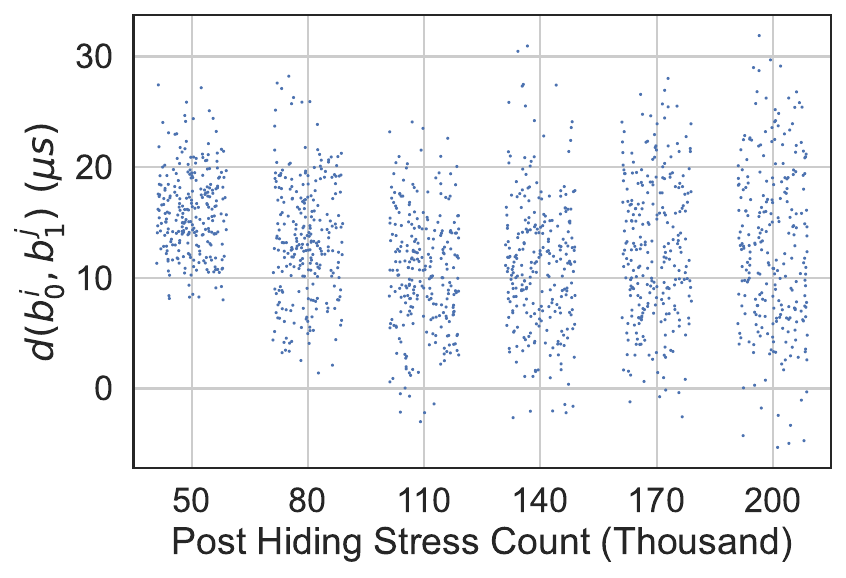}
        \caption{$\mathcal{N}=30K$}
        \label{fig:ResetT_robust_30K}
    \end{subfigure}%
    \vspace{\medskipamount}
    \begin{subfigure}[t]{1\textwidth}
        \centering
        \includegraphics[trim=0.1cm 0.1cm 0.1cm 0.1cm, clip, width = 0.9\textwidth]{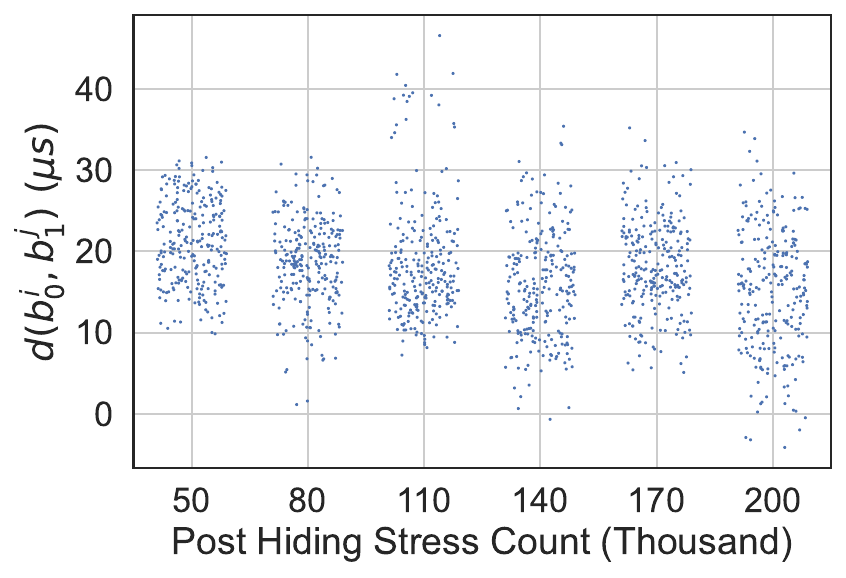}
        \caption{$\mathcal{N}=45K$}
        \label{fig:ResetT_robust_45K}
    \end{subfigure}
\end{minipage}
\begin{minipage} []{.235\textwidth}
    \centering
    \captionsetup{justification=centering, margin= 0cm}
    \begin{subfigure}[t]{1\textwidth}
        \centering
        \includegraphics[trim=0.1cm 0.1cm 0.1cm 0.1cm, clip, width=0.9\textwidth]{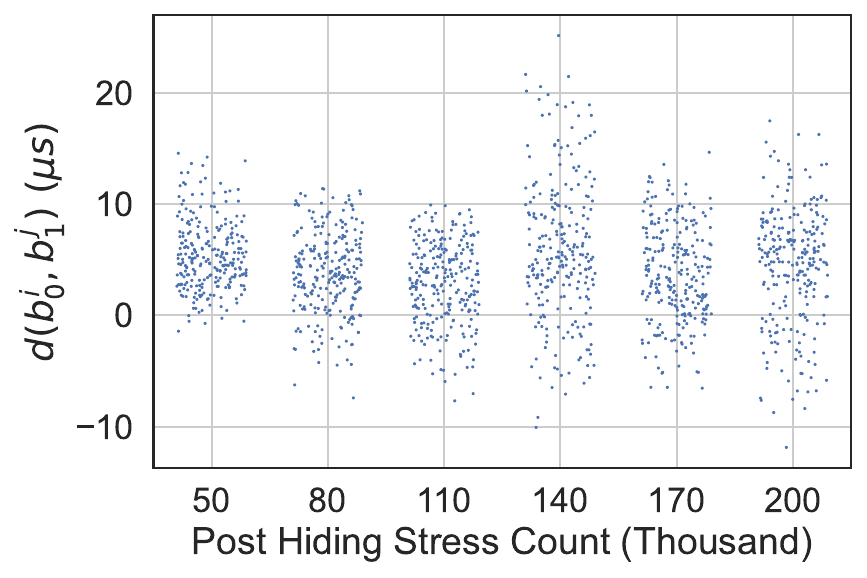}
        \caption{$\mathcal{N}=15K$}
        \label{fig:ResetT_robust_15K_HT}
    \end{subfigure}%
    \vspace{\medskipamount}
    \begin{subfigure}[t]{1\textwidth}
        \centering
        \includegraphics[trim=0.1cm 0.1cm 0.1cm 0.1cm, clip, width = 0.9\textwidth]{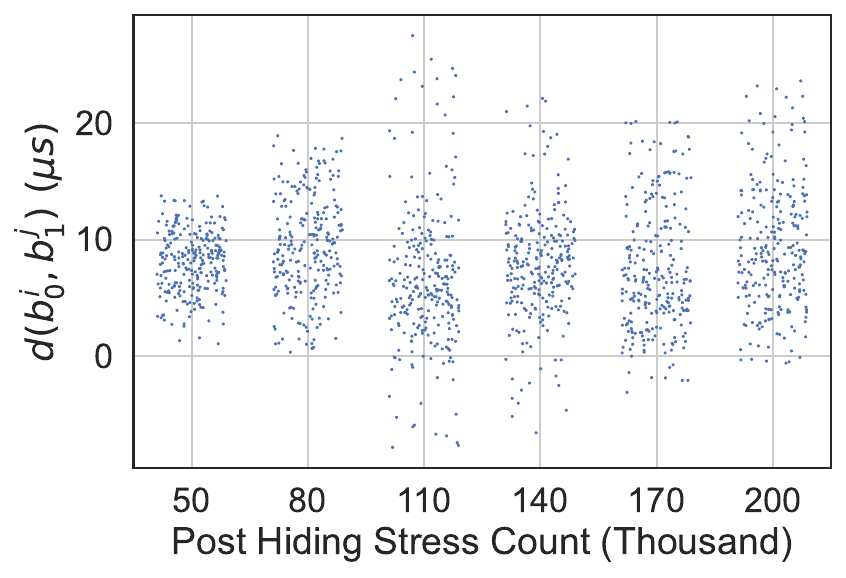}
        \caption{$\mathcal{N}=30K$}
        \label{fig:ResetT_robusts_30K_HT}
    \end{subfigure}%
    \vspace{\medskipamount}
    \begin{subfigure}[t]{1\textwidth}
        \centering
        \includegraphics[trim=0.1cm 0.1cm 0.1cm 0.1cm, clip, width = 0.9\textwidth]{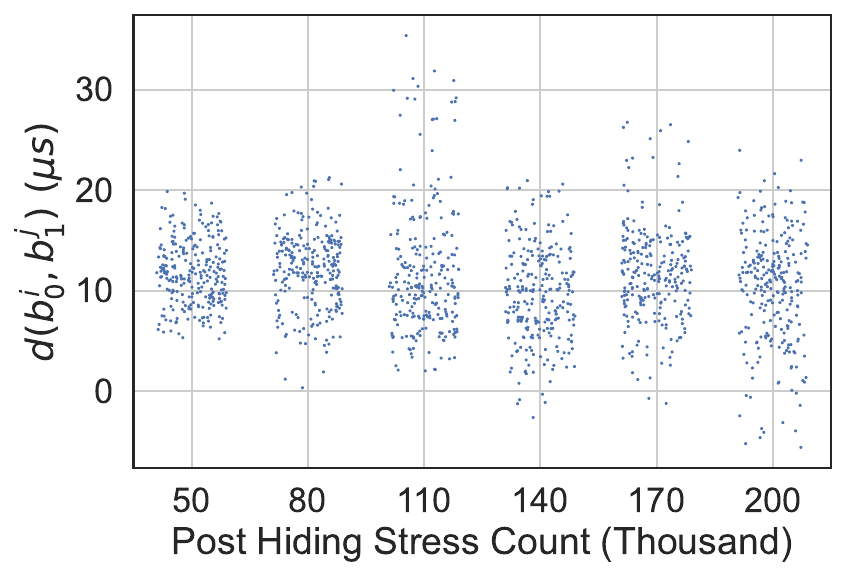}
        \caption{$\mathcal{N}=45K$}
        \label{fig:ResetT_robust_45K_HT}
    \end{subfigure}
\end{minipage}
\caption{Robustness analysis (a)--(c) before (d)--(f) after high-temperature baking ($80^{\circ}C$) with- $t_{Reset,256}$.}
\label{fig:robust_reset}
\end{figure}

\subsection{Security Analysis} \label{subsec:security}

In this section, we perform the security analysis of our proposed technique. The previous section showed how the \textit{write} time is manipulated to store hidden information reliably. Contrarily, in this section, we discuss if an attacker gains access to the memory containing hidden information then whether they can detect hidden information or its existence. 

\subsubsection{Retrieval with inaccurate initial address}

Encoding one bit of secret information in a group of addresses rather than each memory address improves security. If the attacker does not know the hiding key, he or she can not accurately identify the hidden information. Grouping addresses with an inaccurate key will contain both fresh and stressed memory cells; therefore, even the correct threshold value cannot distinguish the fresh and stressed memory cells of the inaccurate group.

Fig. \ref{fig:inac_add} illustrates the retrieved data with an incorrect initial address where the minimum ($\mathcal{N} = 15K$) stress level is used to hide information. The red and blue dots represent the imprinted logic 1s and 0s, respectively. In this experiment, we selected the initial address in three different ways. For example, we used $45K, 30K$, and $15K$ stressing to hide information starting with 32768, 49152, and 65536 memory addresses, respectively. Next, we choose incorrect initial addresses 28672, 32000, and 36864 for $45K$; 45056, 48500, and 53248 for $30K$; and 61440, 65000, and 69632 for $15K$ stress levels, respectively. The reason for choosing the initial addresses in such a way is that \textemdash

\begin{itemize}
    \item {Case 1: The incorrect group overlaps the maximum portion of the correct group, or}
    \item {Case 2: The incorrect initial address resides inside the correct group, or}
    \item {Case 3: The incorrect initial address is an integer multiple of replica size (in our case, replica size is 256).}
\end{itemize}

Although Fig. \ref{fig:inac_add} is constructed with the ($\mathcal{N} = 15K$) stress count, a similar characteristic is valid for a higher stress count (i.e., up to $45K$). Besides, Fig. \ref{fig:inac_add} illustrates case 3 only; a similar characteristic is valid for the other two cases\footnote{If the reviewers want, we will present results for those two cases as well.}. The figure shows that there is no clear separation between the fresh and stressed memory cells. Therefore, the value of hidden bits can not be retrieved through thresholding. This result depicts that it is difficult for attackers to retrieve hidden data without the correct initial address because each group will likely have fresh and stressed memory cells.

\begin{figure}[ht!]
    \centering
    \captionsetup{justification=centering, margin= 0.5cm}
    \begin{subfigure}[t]{0.24\textwidth}
        \centering
        \includegraphics[trim=0.2cm 0.3cm 0.2cm 0.5cm, clip, width = 0.9\textwidth]{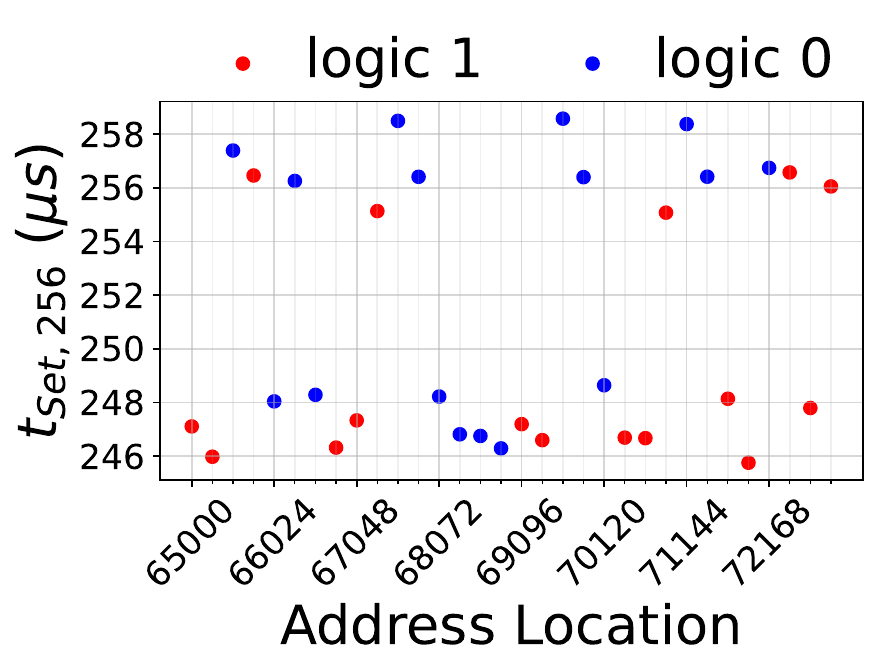}
        \caption{$\mathcal{N}=15K$}
        \label{fig:inac_add_Set}
    \end{subfigure}
    \begin{subfigure}[t]{0.24\textwidth}
        \centering
        \includegraphics[trim=0.3cm 0.3cm 0.2cm 0.5cm, clip, width = 0.9\textwidth]{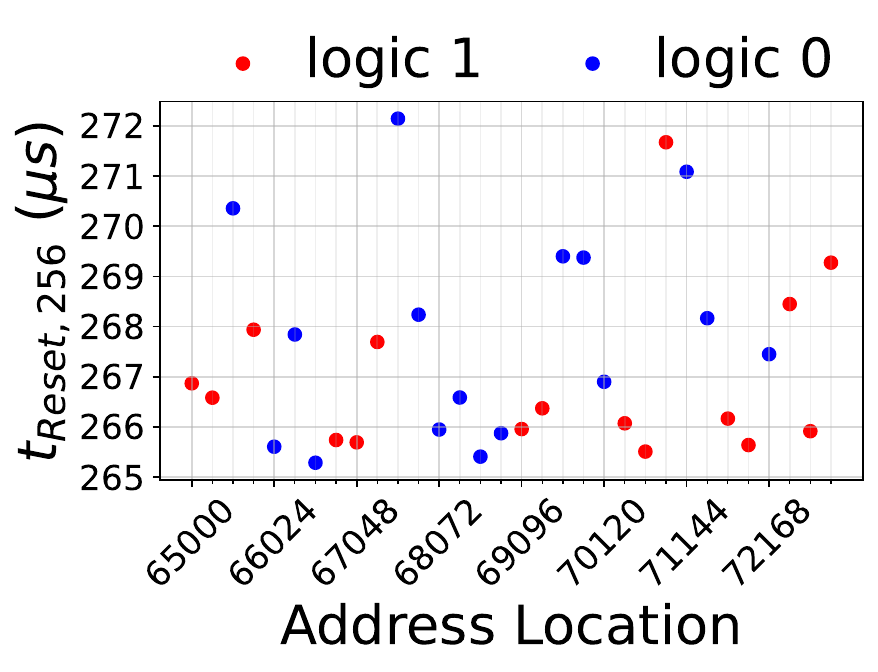}
        \caption{$\mathcal{N}=15K$}
        \label{fig:inac_add_Reset}
    \end{subfigure}
    \caption{Retrieved data with an incorrect initial address at ($15K$) stressing used to hide information with- (a) $t_{Set,256}$ (b) $t_{Reset,256}$.}
    \label{fig:inac_add}
\end{figure}

\subsubsection{Data hide with enhanced security} \label{subsubsec:inac_hideKey}


In the previous sections, we hide \nth{1} bit of hidden data in 256 consecutive addresses, then the \nth{2} bit in another 256 consecutive addresses, and so on. However, the security of our proposed data-hiding scheme can be enhanced in many ways. For example, we can perform the hiding through replication and random rotation to strengthen security. To this end, we hide \nth{1} copy of 32-bit scramble hidden data in 32 consecutive addresses, then the \nth{2} copy in another 32 consecutive addresses, and so on. Fig. \ref{fig:secureDataHide} demonstrates this method using 8-bit data (payload). For simplicity, we assume that replicating the payload only 16 times is sufficient to recover it without any error. However, instead of saving each replica directly to the memory, we rotated (circular shift) each replica with a random displacement ($\mathcal{K}$). The displacement should be uniform within the possible range to maximize security (in this case, $7\leq\mathcal{K}\leq0$). Here, the $\mathcal{K}$ for each replica can be considered as the key. \textcolor{black}{Note that, instead of a circular shift, we can also use the stream cipher algorithm to randomize the pattern \cite{cusick2004stream}.} While retrieving the hidden information, the key is required to reverse the rotation in order to find the appropriate set of \textit{set/reset} time for each information bit. Therefore, the security of the information can be guaranteed as long as the key is uncompromised.

\begin{figure}[ht!]
    \centering
    \captionsetup{justification=centering, margin= 0cm}
    \includegraphics[trim=0cm 1cm 20.5cm 0cm, clip, width = 0.3\textwidth]{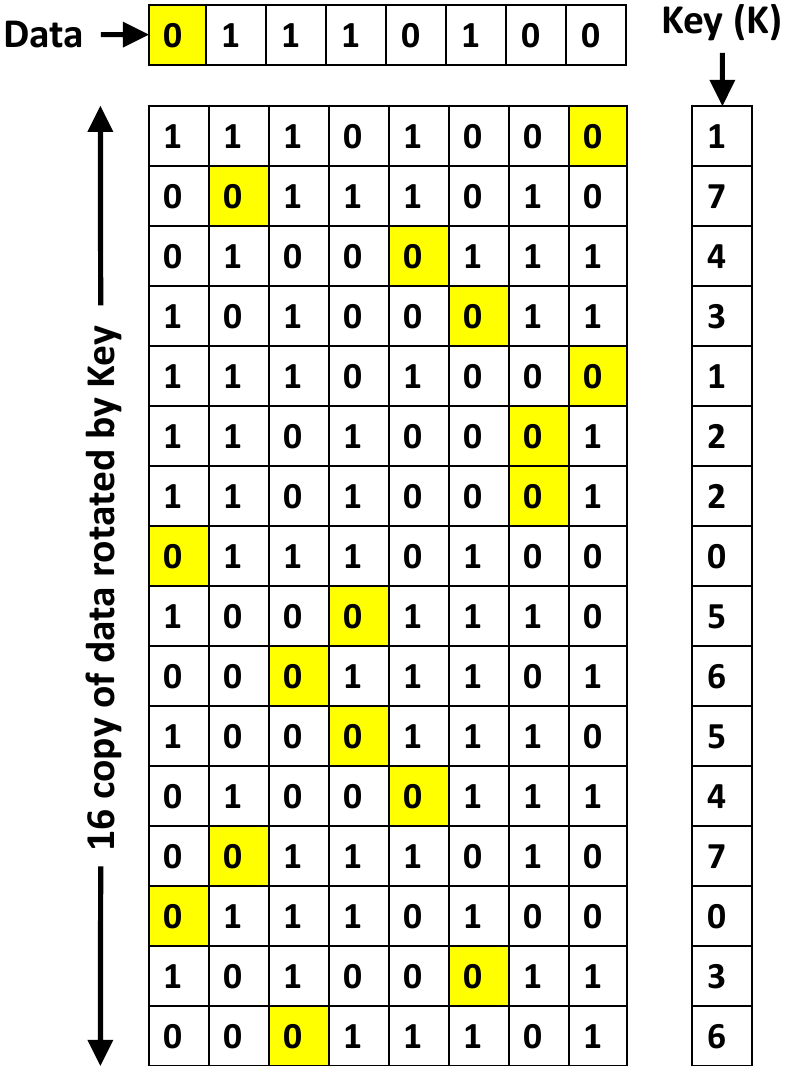}
    \caption{Data hiding with enhanced security: replication + left circular rotation. The start bit is marked with yellow to track the rotation.}
    \label{fig:secureDataHide}
\end{figure}

\begin{figure}[ht!]
    \centering
    \captionsetup{justification=centering, margin= 0.5cm}
    \begin{subfigure}[t]{0.24\textwidth}
        \centering
        \includegraphics[trim=0.2cm 0.3cm 0.2cm 0.5cm, clip, width = 0.9\textwidth]{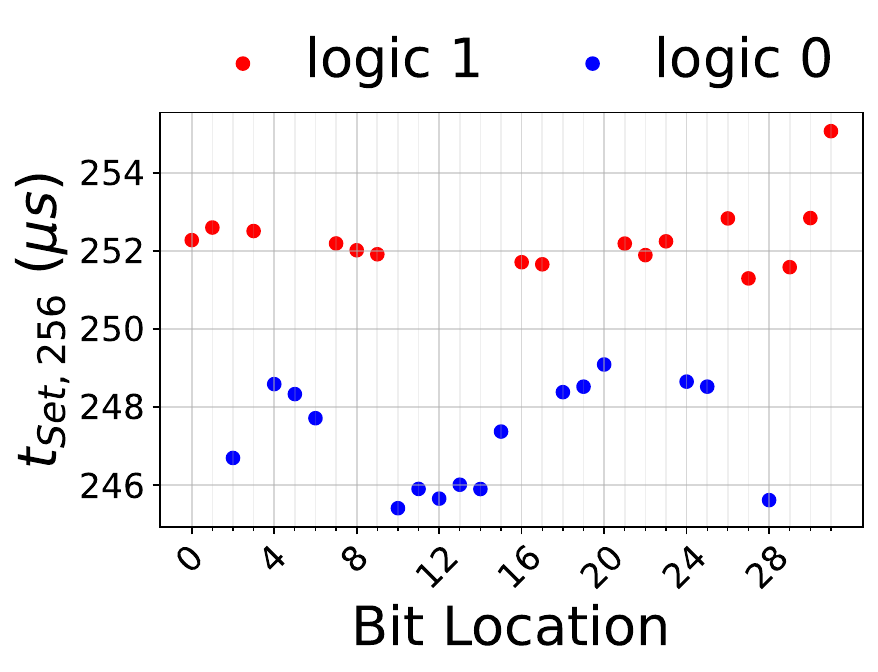}
        \caption{$\mathcal{N}=15K$}
        \label{fig:corr_add_Set}
    \end{subfigure}
    \begin{subfigure}[t]{0.24\textwidth}
        \centering
        \includegraphics[trim=0.3cm 0.3cm 0.2cm 0.5cm, clip, width = 0.9\textwidth]{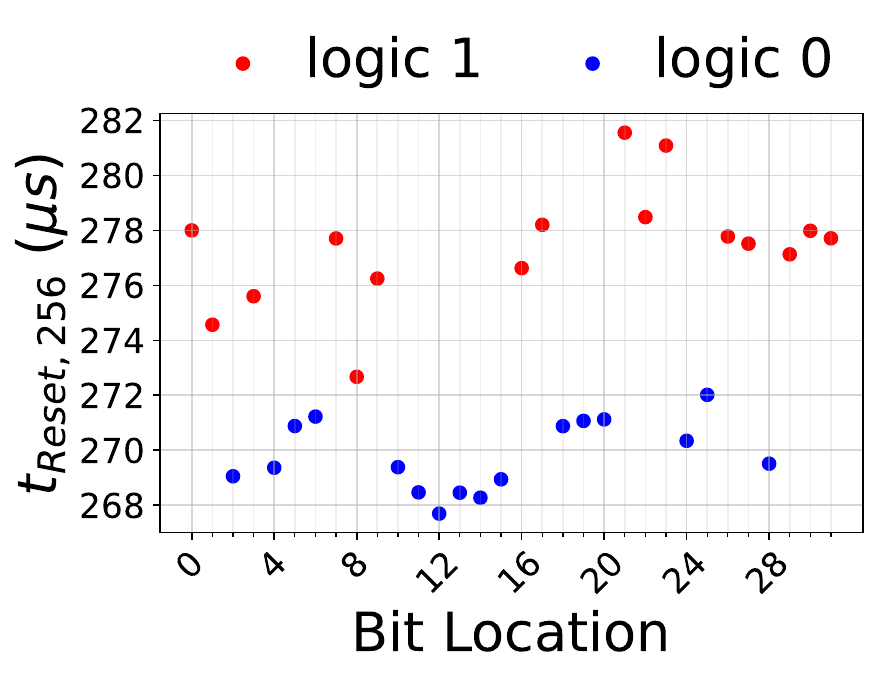}
        \caption{$\mathcal{N}=15K$}
        \label{fig:corr_add_Reset}
    \end{subfigure}
    \caption{Retrieved data with a correct key with $\mathcal{N} = 15K$ stressing used to hide information with- (a) $t_{Set,256}$ (b) $t_{Reset,256}$.}
    \label{fig:corr_key}
\end{figure}

Fig. \ref{fig:corr_key} is constructed with the correct key using a $ \mathcal{N} = 15K$ stress level to hide information (with 256 replicas). A similar characteristic is valid for a higher stress count (i.e., up to $45K$). The figure shows that there is a clear separation between the fresh and stressed memory cells. On the other hand,  Fig. \ref{fig:inac_key} is constructed with an incorrect key with the same $15K$ stress level to hide information. However, there is no clear separation between the fresh and stressed memory cells. Therefore, the value of hidden bits can not be retrieved through thresholding. This result depicts that attackers find it difficult to retrieve hidden data without the correct key because each group will likely have fresh and stressed memory cells.

\begin{figure}[ht!]
    \centering
    \captionsetup{justification=centering, margin= 0.5cm}
    \begin{subfigure}[t]{0.24\textwidth}
        \centering
        \includegraphics[trim=0.2cm 0.3cm 0.2cm 0.5cm, clip, width = 0.9\textwidth]{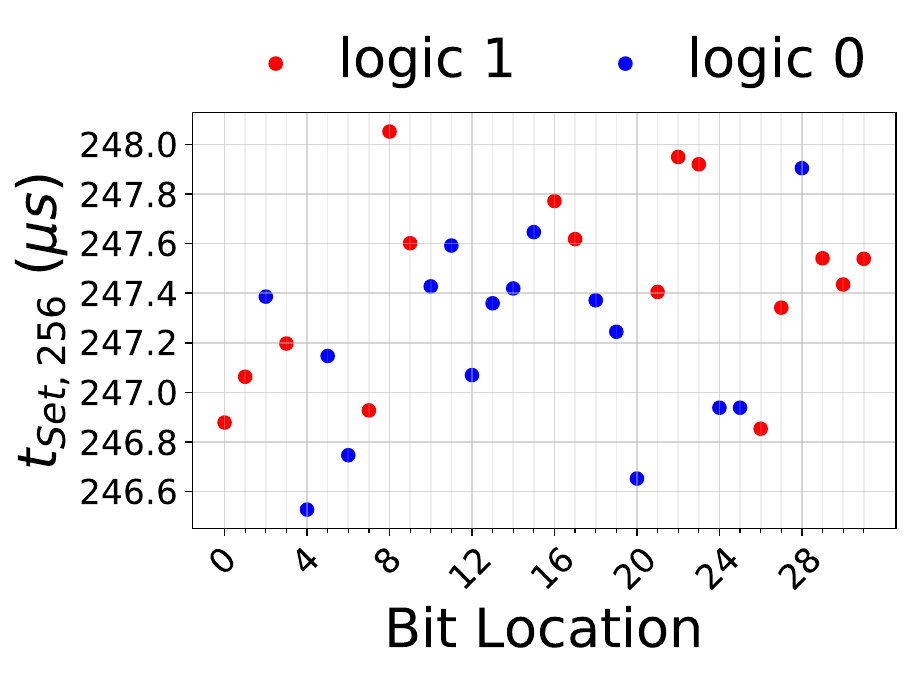}
        \caption{$\mathcal{N}=15K$}
        \label{fig:inac_key_Set}
    \end{subfigure}
    \begin{subfigure}[t]{0.24\textwidth}
        \centering
        \includegraphics[trim=0.3cm 0.3cm 0.2cm 0.5cm, clip, width = 0.9\textwidth]{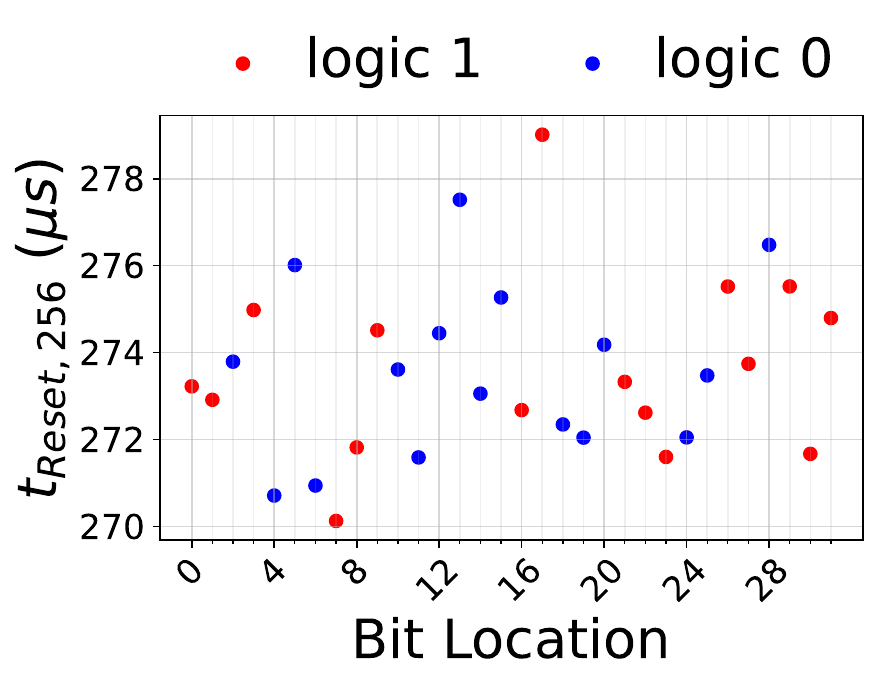}
        \caption{$\mathcal{N}=15K$}
        \label{fig:inac_key_Reset}
    \end{subfigure}
    \caption{Retrieved data with an incorrect key with $\mathcal{N} = 15K$ stressing used to hide information with- (a) $t_{Set,256}$ (b) $t_{Reset,256}$.}
    \label{fig:inac_key}
\end{figure}

However, given an unlimited amount of time, an attacker may extract the hidden information by distinguishing the stress and fresh cells by observing the \textit{write (set/reset)} time of the memory cells. Fortunately, NVM chips usually have millions of cells, and getting timing information for individual cells is very time-consuming and expensive. Moreover, the "hiding key" (Fig. \ref{fig:dataHide_step}) gives another layer of security and is also kept in secure memory and not available to attackers. Furthermore, different ReRAM cells age at different rates since the data written into the memory chip are random. Therefore, with the presence of used cells, an attacker would experience more false positives and false negatives than a fresh ReRAM chip because some cells are less stressed than others, and they are randomly distributed. Finally, since stressed and fresh cells provide different timing information, we can add random delays at the hardware or software level to provide another layer of security. This random timing noise can easily hide the timing information of stressed and fresh cells. The random delay block will be active during the regular operation and inactive while we retrieve the hidden data. We assume that the attacker has no control over this random delay block and requires administrative privilege to deactivate the random delay block during the hidden data recovery mode. \textcolor{black}{As long as the key is uncompromised, the attacker cannot tamper with the proposed security scheme. Therefore, the \textit{probability of coincidence} \cite{PoC_Review4} will remain low.}

\subsection{Potential Application Scenarios }

Given the initial and post-hiding stress tolerance and easy applicability of the proposed scheme on commercial ReRAM chips, we believe the technique can be used in several interesting applications. \textcolor{black}{However, our proposed scheme is not a replacement for traditional encryption but rather more comparable to physical steganography in digital information, i.e., adding new information on top of existing information without any trace (e.g., without requiring additional storage space). Likewise, in steganography, data hiding should not be used as an alternative to data encryption techniques. However, data hiding can still be useful in several applications, such as making robust watermarks (which can be recovered after using memory for a long period of time), second-layer protection over encrypted data, secure and covert data storage, data integrity, covert military/police communication, and digital rights management to protect intellectual property and data from tampering, etc.} For example, one can hide sensitive secret information in the ReRAM-based storage device with higher confidence that others cannot retrieve the information even when the device is lost or stolen. Additionally, an attacker might go after the data to sell it, disrupt operations, seek revenge, or use it for political or competitive advantage. Therefore, data hiding offers an extra protection layer over conventional encryption by thwarting an adversary from reading/copying the ciphertext. Moreover, our proposed data-hiding technique can be extremely useful in some particular attack situations. For example, in a ransomware attack, the attacker modifies the data with some key and demands ransom from the victim in exchange for the key. However, modifying the hidden data with our proposed algorithm will require repetitive  \textit{set/reset} operations and can be easily detected by the system by monitoring the number of \textit{set/reset} operations in a particular address. 

\subsection{Summary of Results}

Overall, we draw the following main conclusions from the results.
\begin{enumerate}[leftmargin=*, topsep=0pt,itemsep=-1ex,partopsep=1ex,parsep=1ex]
    \item {
    The uniform switching characteristics of memory chips sharing the same part-number make it possible to sample a small set of memory chips from each part-number and perform cell characterization over those chips only. Additionally, we only need to characterize once to understand the switching properties.}
    \item {Our silicon result shows that the imprinting and retrieval throughput is $0.4bit/min$ and $15.625bits/s$, respectively. The throughput will be even higher if the hiding scheme uses a smaller replica size.}
    \item {Retention time has little or no impact over bit separation.}
    \item {The \textit{set} operation can endure higher initial and post-hiding stress levels than the \textit{reset} operation.}
    \item {The hidden information is not significantly affected by high-temperature baking and high-temperature system-level operation.}
    \item {The hiding key is a security parameter that can be adjusted to provide greater or lesser protection against brute force attacks.}
\end{enumerate}

\section{Conclusion} \label{sec:end}

This paper demonstrated a cost-effective technique to hide information using the memory cells' \textit{set/reset} time of commercially available ReRAM chips. \textit{Write} (\textit{set/reset}) time of ReRAM is an analog physical characteristic that has no relation with the stored digital content and the normal memory usage. Besides, the stored information can be survived successfully even after thousands of memory operations. In our proposed technique, we utilize repeated switching operations to change the physical properties of the memory cells. Without adequate knowledge about the hiding key, analog physical characteristics measurement will not help reveal the hidden information. The effectiveness of the proposed technique is evaluated by metrics of interest, i.e., the bit separation and hiding cost. Additionally, our proposed technique is robust against temperature variation and does not require any hardware modifications.

\section*{Acknowledgments}
This work was supported by the National Science Foundation under Grant Number DGE-2114200.



\bibliographystyle{IEEEtran}
\bibliography{IEEEabrv,ReRAM_dataHide_TIFS}

 
%

\ifCLASSOPTIONcaptionsoff
  \newpage
\fi

 
\vspace{11pt}

\vspace{-33pt}
\begin{IEEEbiography}[{\includegraphics[width=1in,height=1.25in,clip,keepaspectratio]{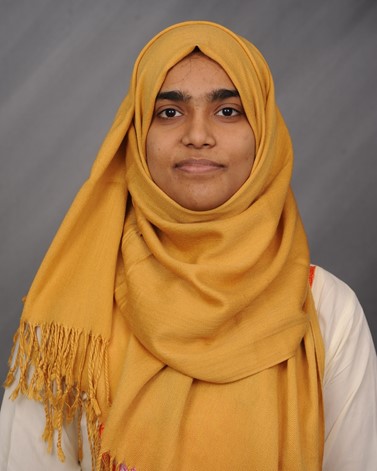}}]{Farah Ferdaus} (S'20–M'24) is a Postdoctoral Researcher at Argonne National Laboratory. She received her Ph.D. degree in Electrical and Computer Engineering from Florida International University in 2022.  
Her research interests include AI Accelerators, high-performance energy-efficient computer architecture, emerging memory technologies, hardware security, and machine-learning applications.
\end{IEEEbiography}
\vspace{-1.2cm}
\vskip 0pt plus -1fil 
\begin{IEEEbiography}
[{\includegraphics[width=1in,height=1.25in,clip,keepaspectratio]{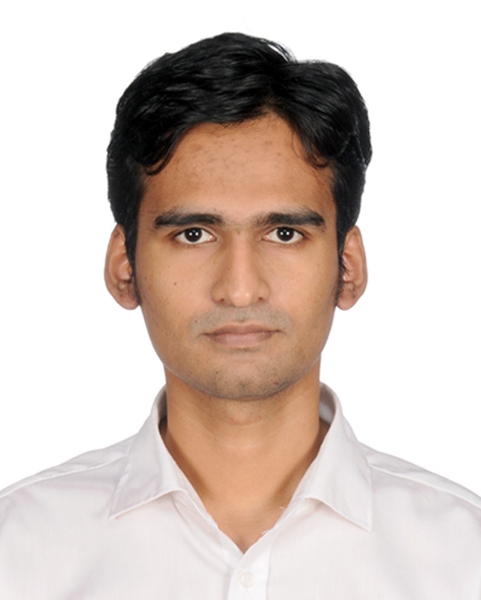}}]
{B. M. S. Bahar Talukder} (S'18) received his Ph.D. degree in Electrical and Computer Engineering from Florida International University. 
His primary research interests include hardware security, secured computer architecture, machine-learning application in system security, and emerging memory technologies.
\end{IEEEbiography}
\vspace{-1.2cm}
\vskip 0pt plus -1fil 
\begin{IEEEbiography}
[{\includegraphics[width=1in,height=1.25in,clip, trim=0cm 0cm 0cm 0cm, keepaspectratio]{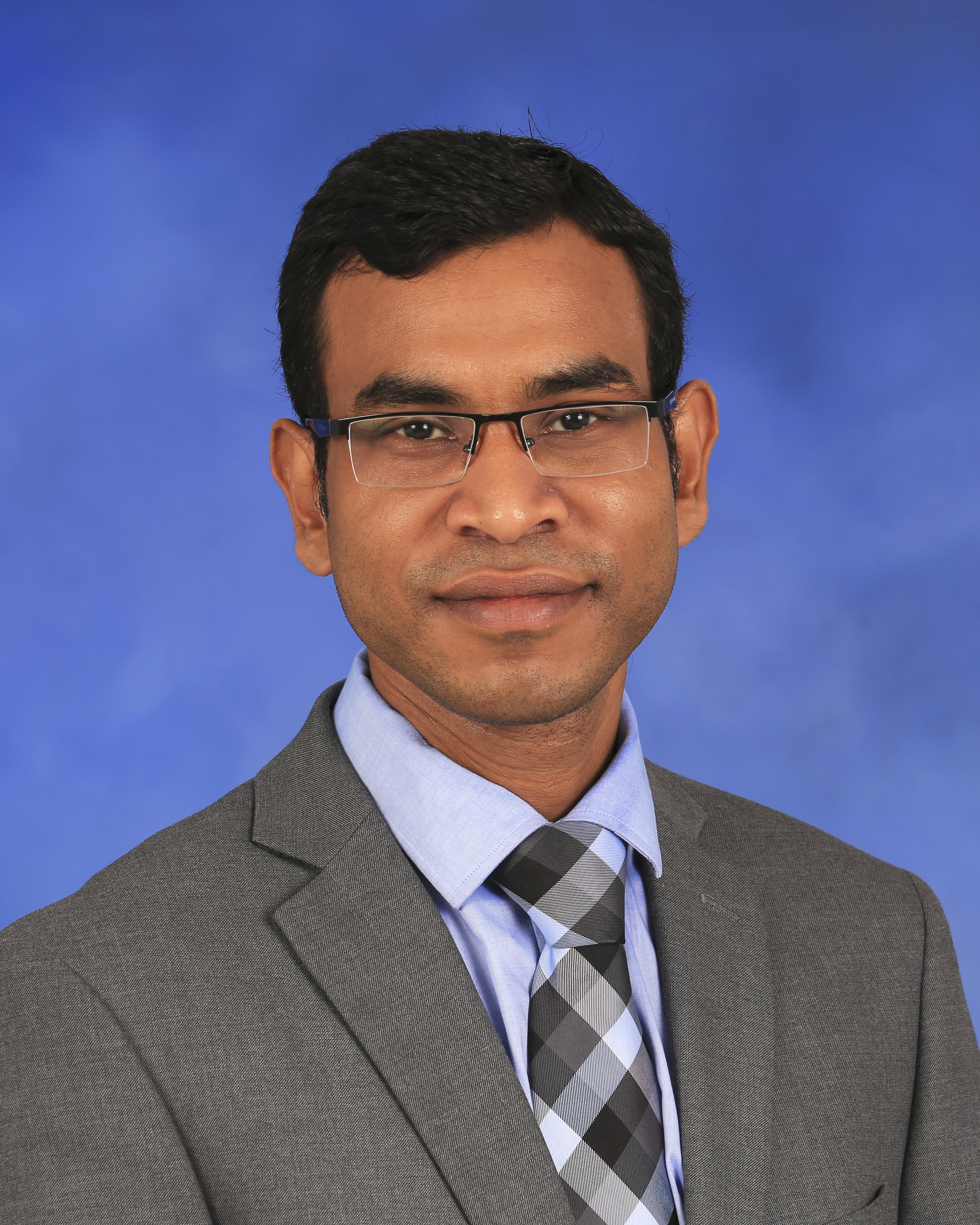}}]
{Md Tauhidur Rahman} (S'12–M'18-SM'21) is an Assistant Professor in the Department of Electrical and Computer Engineering at Florida International University (FIU). He received his Ph.D. degree in Computer Engineering from the University of Florida in 2017. His current research interests include hardware security and trust, memory system, machine learning applications, embedded security, and reliability. He is the recipient of NSF CRII. 
\end{IEEEbiography}

\vspace{11pt}


\vfill

\end{document}